\documentclass[lettersize,journal]{IEEEtran}
\usepackage{amsmath,amsfonts}
\usepackage{algorithm}
\usepackage{algpseudocode}
\usepackage{amsmath}

\usepackage{upgreek}
\usepackage{graphicx}
\usepackage{subfigure}
\usepackage{amsthm,amsmath,amssymb,amsfonts}
\usepackage{amsmath,bm}
\usepackage{mathrsfs,amsmath}
\usepackage{enumerate}
\usepackage{amssymb}
\usepackage{array}
\usepackage{longtable}
\usepackage{booktabs}
\usepackage{makecell}

\usepackage[mathscr]{euscript}
\usepackage[square, comma, sort&compress, numbers]{natbib}
\usepackage{color}
\usepackage{url}

\usepackage{setspace}

\ifCLASSINFOpdf
\else
\fi

\begin{document}

\title{Channel Characterization of IRS-assisted Resonant Beam Communication Systems}

\author{Wen Fang, Wen Chen,~\IEEEmembership{Senior Member,~IEEE}, Qingqing Wu,~\IEEEmembership{Senior Member,~IEEE}, Xusheng Zhu, \\Qiong Wu,~\IEEEmembership{Senior Member,~IEEE}, Nan Cheng,~\IEEEmembership{Senior Member,~IEEE}

\thanks{
W.~Fang, W.~Chen, Q.~Wu, and X.~Zhu are with the Department of Electronic Engineering, Shanghai Jiao Tong University, Shanghai 200240, China (e-mail: wendyfang@sjtu.edu.cn, wenchen@sjtu.edu.cn, qingqingwu@sjtu.edu.cn, xushengzhu@sjtu.edu.cn).

Q.~Wu is with the School of Internet of Things Engineering, Jiangnan University, Wuxi 214122, China (e-mail: qiongwu@jiangnan.edu.cn).

N.~Cheng is with the State Key Lab. of ISN and School of Telecommunications Engineering, Xidian University, Xi’an 710071, China (e-mail: dr.nan.cheng@ieee.org).
}}



\maketitle

\begin{abstract}
To meet the growing demand for data traffic, spectrum-rich optical wireless communication (OWC) has emerged as a key technological driver for the development of 6G. The resonant beam communication (RBC) system, which employs spatially separated laser cavities as the transmitter and receiver, is a high-speed OWC technology capable of self-alignment without tracking. However, its transmission through the air is susceptible to losses caused by obstructions. In this paper, we propose an intelligent reflecting surface (IRS) assisted RBC system with the optical frequency doubling method, where the resonant beam in frequency-fundamental and frequency-doubled is transmitted through both direct line-of-sight (LoS) and IRS-assisted channels to maintain steady-state oscillation and enable communication without echo-interference, respectively. Then, we establish the channel model based on Fresnel diffraction theory under the near-field optical propagation to analyze the transmission loss and frequency-doubled power analytically. Furthermore, communication power can be maximized by dynamically controlling the beam-splitting ratio between the two channels according to the loss levels encountered over air. Numerical results validate that the IRS-assisted channel can compensate for the losses in the obstructed LoS channel and misaligned receivers, ensuring that communication performance reaches an optimal value with dynamic ratio adjustments.
\end{abstract}

\begin{IEEEkeywords}
Optical wireless communication, Resonant beam communication, Intelligent reflecting surface, Near-field Fresnel diffraction
\end{IEEEkeywords}

\section{Introduction}
\IEEEPARstart{T}{he} wireless communication technology has undergone significant transformations over the past few decades, evolving towards $6$G driven by the ever-growing demand for higher data rates, enhanced reliability, and increased capacity~\cite{salih2020evolution}~\cite{dang2020should}. The evolution from earlier radio communication to $5$G and beyond has been marked by remarkable advancements in modulation schemes, hardware capabilities, and spectrum utilization, which have collectively enabled a myriad of applications, ranging from mobile internet access to the Internet of Everything (IoE), connecting billions of devices worldwide~\cite{jain2021metaheuristic}. However, as the number of connected devices continues to surge and data-intensive applications like augmented reality (AR), virtual reality (VR), and high-definition video streaming become more prevalent, the traditional radio frequency (RF) communication systems face escalating challenges, including spectrum scarcity, signal attenuation, and electromagnetic interference~\cite{sharma2020toward}~\cite{liang2021realizing}. To address these limitations, alternative communication technologies are being explored, with optical wireless communication (OWC) emerging as a promising candidate benefiting from the spectrum-rich, spectrum license-free, and electromagnetic interference-free~\cite{uysal2014optical}.

The OWC leverages visible, infrared, and ultraviolet light for data transmission, offering a compelling alternative to traditional RF communication systems due to its high bandwidth, immunity to RF interference, and enhanced security from the confined transmission channels~\cite{garg2023next}. The existing OWC encompasses various technologies including visible light communication (VLC) and free-space optical (FSO) communication, where the former utilizes light-emitting diodes (LEDs) for both illumination and data transmission, modulating light intensity at high speeds to transmit data imperceptible to the human eye~\cite{ismail2020review}. However, VLC data rates decrease rapidly with extended coverage due to diminishing light intensity. FSO communication, on the other hand, uses lasers to establish high-speed point-to-point links over free space, providing gigabit data rates~\cite{kaushal2016optical}. Nevertheless, FSO systems require precise alignment of the transmitter and receiver, making installation and maintenance more challenging and limiting the effective range~\cite{khalighi2014survey}.

To address the above limitations of VLC and FSO, a resonant beam-based OWC system utilizing spatially separated laser cavities as the transmitter and receiver, with retroreflectors installed at both ends, has been proposed in~\cite{liu2016charging}~\cite{xiong2019resonant}, allowing the beam to be collinearly reflected. Firstly, the collinear properties of the retroreflector further facilitate self-alignment without the need for tracking, thus enhancing automatic beamforming. Additionally, due to resonant beams' collimation properties, they exhibit high-power and large-range characteristics like lasers. Besides, in a resonant beam system (RBS), the gain medium in the transmitter (Tx) continuously absorbs the energy from the pump source for amplification and excites the resonant beam, which is then transmitted through space to the receiver (Rx)~\cite{fang2021end}. This transmission over air is affected by various losses, including air loss and diffraction loss. When the gain efficiency of excitation and amplification compensates for these transmission losses, the resonant beam transmission reaches a steady state, allowing stable information transmission through the round-trip oscillation between the Tx and Rx~\cite{fang2021safety}. However, communication will automatically weaken or even be interrupted if the transmission loss increases beyond a certain threshold, particularly when external objects obstruct the RB channel. Consequently, the requirement for a line-of-sight (LoS) link remains a significant limitation for the applicability of RBS, similar to other OWC technologies.

Intelligent reflecting surfaces (IRS) represent a breakthbreakth technology that enhances wireless communication by controlling the propagation environment, providing a low-cost and easily deployable solution for achieving coverage enhancement and eliminating blind spots~\cite{li2021joint}~\cite{wu2021intelligent}~\cite{hua2023secure}~\cite{hua2023intelligent}. An IRS is a planar array of passive elements whose electromagnetic properties can be dynamically adjusted to reflect incident signals in desired directions, which has recently been applied to OWC systems to overcome some inherent limitations~\cite{basharat2021reconfigurable}. For VLC,~\cite{aboagye2022ris} provides a comprehensive tutorial on using IRSs in indoor VLC systems to mitigate signal obstructions. It compares optical RISs with RF-RISs and optical relays, and proposes solutions and research directions for integrating optical RISs with emerging technologies to enhance VLC system performance, including non-orthogonal multiple access (NOMA) and multiple-input multiple-output (MIMO) systems. \cite{aboagye2023liquid} and \cite{aboagye2022design} introduce a novel approach utilizing liquid crystal-based IRS-enabled VLC transmitters and receivers in indoor VLC systems. The former demonstrates superior performance in data rate and illumination compared to traditional VLC transmitters, while the latter enhances signal strength and data rate by steering incident light and amplifying intensity in the LC RIS-based receiver.~\cite{abdelhady2022channel} investigates the temporal characteristics of IRS-based VLC channels, deriving impulse responses and delay spreads for general and specific configurations, and provides exact, bound, and asymptotic delay expressions. Besides, the communication performance including spectral efficiency and sum rate is optimized in \cite{cang2022joint},~\cite{sun2022joint} and \cite{liu2023sum} by the joint resource management method and power-domain NOMA. On the other hand, \cite{najafi2021intelligent} investigates the use of IRSs to mitigate the LoS requirement in FSO systems, designing a phase-shift distribution for IRSs to control beam direction and developing a statistical model for geometric and misalignment losses. \cite{sipani2023modeling} presents a practical approach to model and design an IRS-assisted FSO communication system considering random misalignment, where a geometric and misalignment loss (GML) model that accounts for the statistical distribution of incidence and reflection angles is established for driving the analytical closed-form expressions. A phase-shift design for IRSs to focus beams in FSO systems is proposed in \cite{noh2023phase}, where a new pointing error model and an outage performance analysis with Monte Carlo simulations are conducted. Then, \cite{sikri2021reconfigurable}, \cite{wang2023uplink}, and \cite{sikri2022signal} study the co-channel interference, dual-hop effect, and other communication performance aspects in IRS-assisted FSO-RF systems, demonstrating significant performance enhancement compared to traditional FSO/RF systems.

Therefore, to enhance the communication performance in the absence of LoS channels, we propose using IRS to mitigate transmission losses over the resonant cavity (i.e., the free space between the Tx and Rx) in a resonant beam communication (RBC) system. Meanwhile, to address the issue of echo interference caused by the round-trip propagation of the resonant beam, we adopt an optical frequency doubling method, which maintains the excitation process in the gain medium of the fundamental-frequency beam within the resonant cavity, while using the frequency-doubled beam for information loading and transmission. Moreover, in the RBC transmitter, the transmitted power can be dynamically allocated between the direct LoS channel and the IRS-assisted channel by controlling the beam-splitting ratio based on the loss level over air, thereby ensuring that communication performance is maximized. The main contributions of this work can be summarized as follows.

\begin{itemize}
    \item We propose the IRS-assisted RBC system design incorporating the optical frequency doubling method, where the resonant beam can be transmitted in both direct LoS and IRS-assisted channels. The frequency-doubled beam, generated by second harmonic generation (SHG), is used for downlink communication, while the fundamental-frequency beam is responsible for oscillations.
    \item We establish an IRS-assisted RBC channel model using Fresnel diffraction under near-field optical transmission conditions, where the channel gain is influenced by the effective transfer aperture, transmission distance, and the relative positions of the IRS and receivers.
    \item Based on transmission loss caused by external object obstruction, receiver misalignment, and changes in distance, the beam-splitting ratio in the Tx can be dynamically adapted to maximize communication power. Numerical evaluation reveals that channel capacity of RBC can be maximized using the ratio optimal algorithm.
\end{itemize}

The remainder of this paper is organized as follows. The RBC system model and signal models are presented in Section II. The transmission model in direct LoS and IRS-assisted channels under near-filed propagation is established in Section III. The power for the frequency-fundamental beam and frequency-doubled beam, the channel capacity of the IRS-assisted RBC system are derived in Section IV. Simulation results of communication performance with different parameters are demonstrated in Section V. Finally, the conclusions are drawn in Section VI.

\begin{figure*}[!t]
\centering
\includegraphics[width=6.0in]{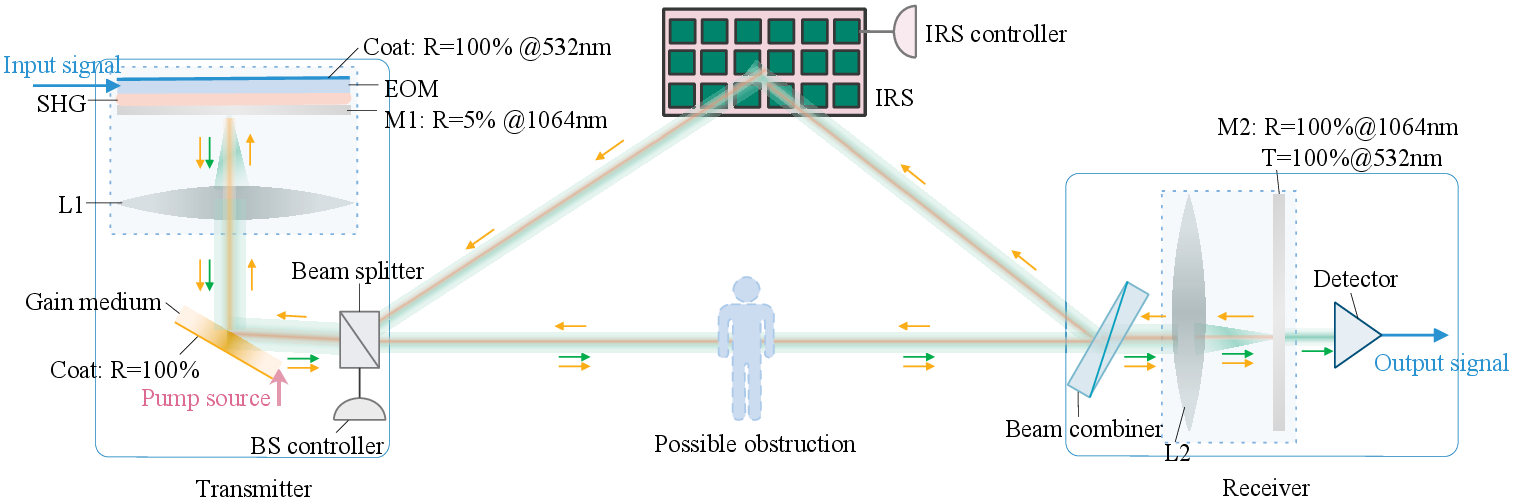}
\caption{The example system structure of IRS-assisted RBC system.}
\label{fig_1}
\vspace{-10pt}
\end{figure*}

\section{System Model}
In this section, we first introduce the system structure and work process of the IRS-assisted RBC system. Then, we present the near-field characteristics of resonant beam propagation. Next, we establish the signal model for resonant beam propagation, considering free-space optical transmission.

\subsection{IRS-assisted RBC system}
The system structure of the IRS-assisted RBC system, depicted in Fig.~\ref{fig_1}, encompasses a spatially separated transmitter and receiver, constituting a resonant cavity for free space transmission. The input reflector including plane mirror M$1$ and confocal lens L$1$, second harmonic generation (SHG), electro-optic modulator (EOM), beam splitter, and controller make up the transmitter, while the beam combiner, output reflector including confocal lens L$2$ and plane mirror M$2$, and detector are comprised of the receiver. The photons, amplified by the pump source in the gain medium, propagate forth and back through free space, reflect off the output reflector, and then undergo another free-space transmission back to the transmitter. Moreover, the amplification in the gain medium compensates for the transfer loss in the resonant cavity, enabling the system to reach a stable state and forming the standing wave.

Additionally, the SHG crystal serves as an electromagnetic wave frequency-doubling device that takes in a fundamental frequency beam and generates a second frequency-doubled beam as output. For instance, the SHG crystal is precisely embedded in M$1$, facilitating the doubling of the $1064$ nm resonance beam to a $532$ nm beam. After the completion of information loading by the EOM, the frequency-doubled beam is transmitted to the receiver via the optical channel of the fundamental beam for communication. Furthermore, reflective coatings with varying reflectivity for different frequency bands are applied to M$1$, EOM, and the gain medium to facilitate the reflection and transmission of beams at both fundamental and doubled frequencies. In the receiver, the output reflector selectively outputs the frequency-doubled beam, which is photoelectrically converted in the detector to transfer the signal. Due to the co-linear reflection characteristics of retro-reflectors, the resonant beam can establish self-alignment and achieve a stable state.

According to the above introduction, if an external object blocks the LoS optical channel, the stable state for both fundamental and doubled frequency beams can no longer be maintained due to the sharply increased loss, resulting in a degradation in communication performance. To ensure optimal performance, we propose adopting an IRS to adjust the optical phase for NLoS transmission in free space. As illustrated in Fig.~\ref{fig_1}, if a possible obstruction occurs in the LoS transmission, the beam controller transmits a portion of the resonant beam in the LoS channel and the remainder in the IRS-assisted channel via the beam splitter.

In summary, take Fig.~\ref{fig_1} as an example, the work process of IRS-assisted RBC system can be expressed by the following principles: i) the resonant beam at fundamental frequency, i.e. wavelength at $1064$ nm, propagates back and forth in the resonant cavity; ii) $95\%$ of the fundamental-frequency beam passes through M$1$ and then through SHG to form doubled-frequency beam, i.e. wavelength at $532$ nm; iii) EOM loads information onto the doubled-frequency beam; iv) the beams at fundamental and doubled frequencies transfer through the direct LoS and IRS-assisted channels to receiver under the control of beam splitter in a certain ratio; v) the output reflector in the receiver outputs the doubled-frequency beam and reflects the fundamental-frequency beam to the transmitter; and vi) the detector performs photoelectric conversion and completes the downlink information transmission. Moreover, the steady-state and retro-reflective properties of the resonant cavity allow the resonant beam to maintain radiation safety by blocking interruptions and achieve mobility through self-tracking.

\vspace{-5pt}
\subsection{Near Field Characteristic}
\begin{figure}[!t]
\centering
\includegraphics[width=3.5in]{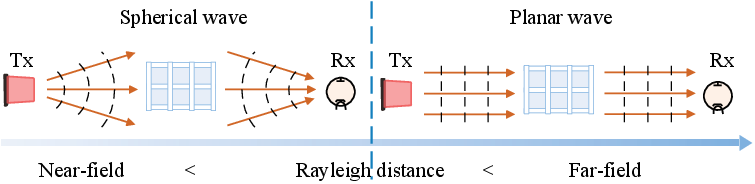}
\caption{The schematic of the near-field and far-field according to Rayleigh distance.}
\label{fig_Rayleigh}
\vspace{-10pt}
\end{figure}

As a type of electromagnetic field, the resonant beam exhibits both near-field and far-field propagation behaviors, which are manifested in phase difference. In far-field propagation, the resonant beam is modeled as a planar wave, whereas in near-field propagation, it is modeled as a spherical wave~\cite{hao2023far}. The Rayleigh distance is used to demarcate the near-field from the far-field, and it can be expressed as
\begin{equation}\label{eq_raydis}
D_{\mathrm{r}} = \frac{2D^2}{\lambda}.
\end{equation}
where $D$ is the diameter of transfer aperture, and $\lambda$ is the wavelength of transfer beam, which can be calculated by $\lambda=u / v$ with $u$ and $v$ representing the light speed and operating frequency. As shown in Fig.~\ref{fig_Rayleigh}, when the transfer distance between the Tx/Rx and the IRS far exceeds $D_{\mathrm{r}}$, the transfer beam exhibits the propagation specificities of planar wave~\cite{sarieddeen2020next}. Conversely, the near-field property is considered if the transfer distance is less than $D_{\mathrm{r}}$~\cite{cui2022near}.

It is evident that the Rayleigh distance increases as the beam wavelength decreases. Consequently, if the beam wavelength $\lambda$ is sufficiently small with a specific operating aperture, near-field propagation occurs more frequently than far-field propagation, as observed in OWC. For instance, in the case of a RBC system, if the transfer aperture of retro-reflectors in Tx/Rx is $2.5$ mm, the Rayleigh distance for the fundamental-frequency beam (i.e., wavelength at $1064$ nm) is approximately $47$ m, while for the doubled-frequency beam (i.e., wavelength at $532$ nm), it is approximately $93$ m. Given that the application scenarios for the IRS-assisted RBC system typically involve indoor environments, priority is given to the performance of near-field communication.

\subsection{Signal Model}
In this subsection, we develop the signal model of the IRS-assisted RBC system in both direct LoS and IRS-assisted channels. We denote $\mathbf{H}_{\rm{los}}$ and $\mathbf{H}_{\rm{irs}}$ to specify the direct and IRS-reflected LoS channels between Tx and Rx, respectively. Additionally, due to the high transmission speed of the resonant beam, the hardware devices in the RBC system can hardly discern the delays of the multibeam paths, rendering RBC channels frequency-flat.

In general, OWC channels are influenced by transfer losses, misalignment losses, and atmospheric turbulence~\cite{ajam2021channel}. However, in common application scenarios for RBC systems, which are predominantly indoor environments with meter-level distances, atmospheric turbulence can be disregarded. Then, the channel function $\mathbf{H}$ can be modeled as $\mathbf{H}=\mathbf{H}_{\rm{p}}\mathbf{H}_{\rm{g}}$, where $\mathbf{H}_{\rm{g}}$ represents the misalignment losses. Benefiting from the use of retro-reflectors in transceivers, the RBC system can achieve perfect beam tracking. Thus, the positions of the transmitter, IRS, and receiver remain stable at specific locations and the misalignment loss in transmission is mainly caused by the movement of Rx. Besides, $\mathbf{H}_{\rm{p}}$, which indicates the propagation function over air, is characterized by two stages in an IRS-assisted channel, namely Tx-IRS and IRS-Rx, whereas in a direct LoS channel, it is simply represented by Tx-Rx. Thus, in an IRS-assisted channel, $\mathbf{H}_{\rm{p}}$ can be expressed as $\mathbf{H}_{\rm{ir}}\mathbf{H}_{\rm{r}}\mathbf{H}_{\rm{ti}}$, where $\mathbf{H}_{\rm{ti}}$ and $\mathbf{H}_{\rm{ir}}$ denote the channel functions for Tx-IRS and IRS-Rx, respectively, and $\mathbf{H}_{\rm{r}}$ represents the transfer function within the IRS. The transfer function $\mathbf{H}_{\rm{r}}$ can be substituted by $\boldsymbol{\Phi} \in \mathbb{C}^{N\times M}$, where $[\Phi]{n,m} = m e^{j\phi{n,m}}$, with $m$ representing the amplitude reflection, which accounts for absorption and scattering, and $\phi_{n,m}$ being the phase shift argument of the element $(n, m)$.


\section{IRS-Assisted Transmission Model}
In this section, we first determine the power density distribution of the resonant beam based on near-field lightwave propagation. Following this, we investigate the transfer model of both direct LoS and IRS-assisted channels to understand how the IRS influences the propagation of the resonant beam.

\vspace{-5pt}
\subsection{Near-field Propagation Model}
To determine the impact of free-space propagation and the IRS on the resonant beam, we employ scalar field theory to derive the changes in density distribution when the diffracting aperture and transfer distance are larger than the wavelength. As the resonant beam propagates from point $\dot{S}$ in the wave-source plane to point $\dot{P}$ in the observation plane, the wave distribution at point $\dot{P}$ results from a sub-wave that forms after the wave at point $\dot{S}$ reaches point $\dot{Q}$ and is then emitted towards point $\dot{P}$. This propagation process can be accurately described by Kirchhoff's diffraction formula
\begin{equation}\label{eq_Fernel-formal}
\centering
\tilde{U}(P)=\frac{A}{i \lambda} \iint_{\Sigma} \frac{e^{i k l}}{l} \frac{e^{i k r}}{r}\left[\frac{\cos (\vec{n} \cdot \vec{r})-\cos (\vec{n} \cdot \vec{l})}{2}\right] d \sigma,
\end{equation}
where $l$ and $r$ represent the transfer distances between points $\dot{S}$, $\dot{Q}$, and $\dot{P}$, respectively. The wave-number $k$ is defined as $k=2\pi/\lambda$, where $\lambda$ is the wavelength. The symbol $\Sigma$ denotes the set of all sub-waves. For a resonant beam with paraxial propagation, the wave satisfies
\begin{equation}\label{eq_Para-approxi}
\centering
\begin{cases}
r=z_1 \\
\frac{A e^{i k l}}{l} \rightarrow U\left(x_1, y_1\right)\\
\cos (\vec{n} \cdot \vec{r})=1\\
\cos (\vec{n} \cdot \vec{l})=-1
\end{cases},
\end{equation}
Thus, the diffraction transfer formula can be transformed as
\begin{equation}\label{eq_Fernel-trans}
\centering
\tilde{U}(P)=\frac{1}{i \lambda z_1} \iint_{\Sigma} U\left(x_1, y_1\right) e^{i k r} d \sigma,
\end{equation}
where $U\left(x_1, y_1\right)$ is the density distribution on the source aperture. $kr$ represents the phases of spherical sources with $r$ denoting the transfer distance from source point to observe point. For near-field lightwave propagation, the Fresnel approximation can be used to deduce $r$,
\begin{equation}\label{eq_fresnel-appro1}
\centering
r = \sqrt{z_1^2+\left(x-x_1\right)^2+\left(y-y_1\right)^2},
\end{equation}
In addition, the distance of propagation beam axis $z_1$ is larger than transfer aperture, i.e. $z_1 \gg\left(x-x_1\right)^2+\left(y-y_1\right)^2$, the binomial expansion of \eqref{eq_fresnel-appro1} is
\begin{equation}\label{eq_fresnel-appro2}
\begin{aligned}
r &= z_1+\frac{1}{2 z_1}\left[\left(x-x_1\right)^2+\left(y-y_1\right)^2\right]-\frac{1}{8 z_1^3}\\
& \left[\left(x-x_1\right)^2+\left(y-y_1\right)^2\right]^2+\ldots.
\end{aligned}
\end{equation}
When $z_1$ is significantly larger, the higher-order terms can be ignored. Thus, the transfer distance $r$ can be expressed as:
\begin{equation}\label{eq_fresnel-appro3}
\centering
r \approx z_1+\frac{1}{2 z_1}\left[\left(x-x_1\right)^2+\left(y-y_1\right)^2\right].
\end{equation}

Thus, the density at any point $P$, whose coordinates are $(x,y)$ (where the surface perpendicular to the optical axis and parallel to the transfer aperture is the $xoy$ surface), on the observation aperture can be determined using the Fresnel-Kirchhoff formula for near-field propagation:
\begin{equation}\label{eq_f-k-formal}
\begin{aligned}
    \tilde{U}(P)&=\tilde{U}(x, y) \\
     &=\frac{e^{i k z_1}}{i \lambda z_1} \iint_{\Sigma} \tilde{U}\left(x_1, y_1\right) e^{\frac{i k}{2 z_1}\left[\left(x-x_1\right)^2+\left(y-y_1\right)^2\right]} d x_1 d y_1 \\
\end{aligned}
\end{equation}
where $\tilde{U}\left(x_1, y_1\right)$ represents the density distribution on the source aperture. Furthermore, \eqref{eq_f-k-formal} can be written as the convolution of two-dimensional functions
\begin{equation}\label{eq_con}
\centering
U(x, y)= \iint U\left(x_{1}, y_{1}\right) h\left(x-x_{1}, y-y_{1}\right) \mathrm{d} x_{1} \mathrm{d} y_{1}
\end{equation}
and
\begin{equation}\label{eq_con-h}
\centering
h(x, y)=\frac{e^{i k z}}{i \lambda z} e^{\frac{i k}{2 z}\left(x^{2}+y^{2}\right)}.
\end{equation}
Therefore, the diffraction propagation formula for resonant beam in convolution reads
\begin{equation}\label{eq_con-diffraction}
\centering
U(x,y)=U(x_1,y_1)\ast h(x,y),
\end{equation}
where $(x_1,y_1)$ and $(x,y)$ denote the coordinates of the propagation point on the source and observation apertures.

Furthermore, the density distribution on the observation aperture can be obtained through Fourier and inverse Fourier transformations
\begin{equation}\label{f-k-formal}
\begin{aligned}
    U(x, y)&=\mathcal{H}[U(x,y)]\\
&=\mathscr{F}^{-1}\{\mathscr{F}\{U\left(x_1, y_1\right)\} \cdot \mathscr{F}\{h(x, y)\}\}.
\end{aligned}
\end{equation}
The Fourier transformed transfer function is given by
\begin{equation}\label{eq_fieldpro}
\begin{aligned}
H(\nu_x, \nu_y, z) = \mathscr{F}\{h(x, y)\} = \exp \left(i 2 \pi z \sqrt{\frac{1}{\lambda^2}-\nu_x^2-\nu_y^2}\right),
\end{aligned}
\end{equation}
where $\nu_x$ and $\nu_y$ represent the spatial frequency coordinates, and $z$ is the transfer distance between the wave-source plane and the observation plane.


\subsection{Direct LoS Channel Model}
In this subsection, we establish the channel model for RB direct LoS propagation, and the misalignment loss due to receiver mobility is derived from the changes in density distribution.
\subsubsection{Channel Model}
\
\newline
\indent In both Rx and Tx, the resonant beam must propagate through retro-reflectors, with TX also passing through a gain medium. For the input/output retro-reflectors, the interval between the plane mirrors (M$1$ or M$2$) and the convex lenses (L$1$ or L$2$) is denoted as $l_{\rm{c}}$. To ensure maximum pumping efficiency, the gain medium is located at the focal plane of the input reflector, thus they are separated by $f_{\rm{c}}$. The propagation in space of M$1$/M$2$-L$1$/L$2$, L$1$-gain medium can be derived from \eqref{f-k-formal} and \eqref{eq_fieldpro}
\begin{equation}\label{eq_lm}
\begin{aligned}
    &\mathcal{H}_{l_{\rm{c}}}[U(x,y), l_{\rm{c}}]=\mathscr{F}^{-1}\{\mathscr{F}\{U\left(x, y\right)\} \cdot H(v_{x}, v_{y}, l_{\rm{c}})\} \\
    &\mathcal{H}_{f_{\rm{c}}}[U(x,y), f_{\rm{c}}]=\mathscr{F}^{-1}\{\mathscr{F}\{U\left(x, y\right)\} \cdot H(v_{x}, v_{y}, f_{\rm{c}})\}
\end{aligned},
\end{equation}
where $U\left(x, y\right)$ represents the density distribution of the incident resonant beam.

Moreover, the field changes when the incident beam propagates through the apertures, including M$1$, M$2$ L$1$, L$2$, and the gain medium, can be expressed as:
\begin{equation}\label{eq_mirror}
\begin{aligned}
    &\mathcal{P}_{M}[U(x,y), M]=U(x,y) \cdot \begin{cases}
        1, x^2+y^2 \leq r_{\rm{M}}^2\\
        0, else
    \end{cases} \\
    &\mathcal{P}_{L}[U(x,y), L]=U(x,y) \cdot \begin{cases}
        e^{\frac{-i\pi}{\lambda f_{\rm{c}}}(x^2+y^2)}, x^2+y^2 \leq r_{\rm{L}}^2\\
        0, else
    \end{cases} \\
    &\mathcal{P}_{G}[U(x,y), G]=U(x,y) \cdot \begin{cases}
        1, x^2+y^2 \leq r_{\rm{G}}^2\\
        0, else
    \end{cases}
\end{aligned}
\end{equation}
$r_{\rm{M}}$, $r_{\rm{L}}$, and $r_{\rm{G}}$ represent the radii of M$1$/M$2$, L$1$/L$2$, and the gain medium, respectively. The significance of the above equation lies in the fact that the region within the aperture is the effective transmission region, and the action of the focusing mirror $M$ on the optical field is related to its focal length. Hence, the incident resonant beam propagation in transmitter is $\mathcal{T}_{T}\triangleq\mathcal{P}_{G}\mathcal{H}_{f_{\rm{c}}}\mathcal{P}_{L}\mathcal{H}_{l_{\rm{c}}}\mathcal{P}_{M}$, while the propagation in the output reflectors can be defined as $\mathcal{T}_{R}\triangleq\mathcal{P}_{L}^{-1}\mathcal{H}_{l_{\rm{c}}}^{-1}\mathcal{P}_{M}\mathcal{H}_{l_{\rm{c}}}\mathcal{P}_{L}$ with reverse transmission function $(\cdot)^{-1}$. Furthermore, the propagation in free space between Tx and Rx is
\begin{equation}\label{eq_space}
\begin{aligned}
    \mathcal{H}_{D}[U(x,y), D]=\mathscr{F}^{-1}\{\mathscr{F}\{U\left(x, y\right)\} \cdot H(v_{x}, v_{y}, D)\}
\end{aligned}
\end{equation}
where $D$ is the transmission distance. Thus, the direct LoS channel function for resonant beam propagation over a round trip is
\begin{equation}\label{eq_Losround}
\begin{aligned}
\mathbf{H}_{\rm{los}}=\mathcal{T}_{T}^{-1}\mathcal{H}_{D}^{-1}\mathcal{T}_{R}\mathcal{H}_{D}\mathcal{T}_{T}.
\end{aligned}
\end{equation}

If an external object obstructs the direct LoS transmission, it is equivalent to introducing a transfer aperture of decreasing size at the object's location in free space. The impact of external object can be expressed as
\begin{equation}\label{eq_object}
\begin{aligned}
    &\mathcal{P}_{B}[U(x,y), B] =
    \\&U(x,y) \cdot \begin{cases}
        1, x^2+y^2 \leq r_{\rm{B}}^2\  \&\& \ x\ |\ y\leq r_{\rm{B}}-d\\
        0, x^2+y^2 > r_{\rm{B}}^2 \ ||\  x\ |\ y > r_{\rm{B}}-d
    \end{cases},
\end{aligned}
\end{equation}
where $r_{\rm{B}}$ represents the radius of the object aperture, which is always set equal to $r_{\rm{M}}$ or $r_{\rm{L}}$, and $d$ denotes the obstruction depth. The piecewise function in \eqref{eq_object} indicates that the unobstructed area is the effective transmission aperture, expressed as $1$. Based on this premise, the free space propagation can be calculated as $\mathcal{H}_{D} = \mathcal{H}_{D2}\mathcal{P}_{B}\mathcal{H}_{D1}$, where $D1$ and $D2$ represent the distances in Tx-object and object-Rx, respectively. Consequently, the direct LoS channel model with an external object can also be derived in \eqref{eq_Losround}.

\subsubsection{Misalignment Transmission}
\
\newline
\indent Denoting the plane input reflector located on the $xoy$-plane, with the director along the optical axis perpendicular to the plane being the $z$-axis. When the transceivers are strictly aligned, i.e., the output reflector plane is perpendicular to the optical axis, the changes in the density distribution of direct LoS transmission can be calculated using \eqref{eq_Losround}. However, the movement of Rx (translations along the $xoy$ plane and rotations about the axes) will result in misalignment loss.

\begin{figure}[t]
\centering
\subfigure[Translation along $xoy$ plane.]{
\begin{minipage}[t]{\linewidth}
\centering
\includegraphics[scale=0.45]{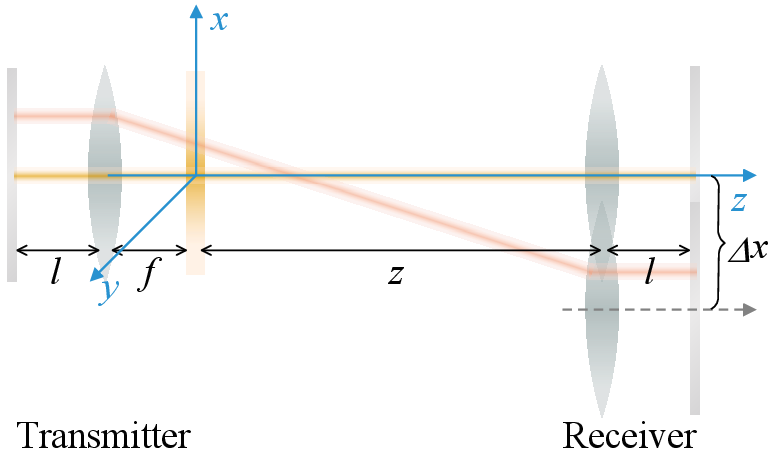}
\label{fig:lostrans}
\end{minipage}
}
\\
\subfigure[Rotation around $x/y/z$-axis.]{
\begin{minipage}[t]{\linewidth}
\centering
\includegraphics[scale=0.45]{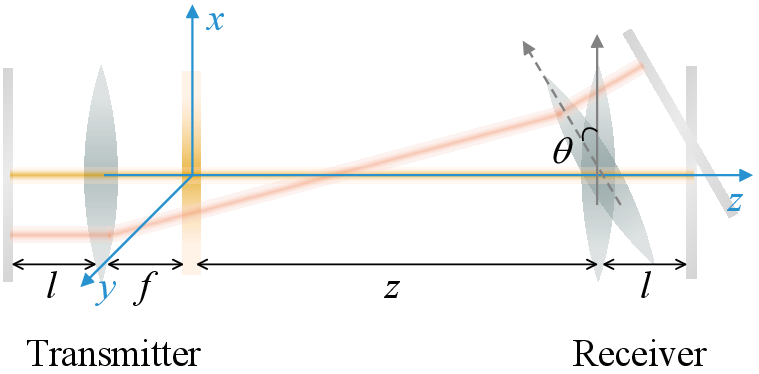}
\label{fig:losrao}
\end{minipage}
}
\centering
\caption{The schematic illustration of LoS channel with misalignment receivers.}
\label{Fig_3}
\vspace{-10pt}
\end{figure}

Firstly, the transformation relationship before and after translation along the $x$, $y$, or $z$ axes is:
\begin{equation}\label{eq_translation}
(x',y',z', 1)^T = M_{\mathrm{t}}(x, y, z,1)^T ,
\end{equation}
where $(x, y, z)$ and $(x',y',z')$ represent the coordinates before and after translation. $M_{\mathrm{t}}$ is the translation matrix,
\begin{center}
    $M_{\mathrm{t}} =\left[\begin{array}{cccc}
    {1} & {0} & {0} & {\Delta x}\\
    {0} & {1} & {0} & {\Delta y}\\
    {0} & {0} & {1}& {\Delta z} \\
    {0} & {0} & {0} & {1}\\
    \end{array}\right],$
\end{center}
$\Delta x$, $\Delta y$ and $\Delta z$ represent the translation distances along the $x$, $y$, and $z$ axes, respectively. On the one hand, when the receiver moves along the $z$-axis, the propagation channel can be calculated using \eqref{eq_space}, with $D$ replaced by $D+\Delta z$. On the other hand, if the receiver translates along the $xoy$ plane as depicted in Fig.~\ref{fig:lostrans}, the misalignment transfer function must incorporate the additional parameters $\exp[i2\pi(v_x \Delta x+v_y \Delta y)]$. Hence, the function of a round-trip transmission with such translation can be expressed as
\begin{equation}\label{eq_transfun}
\begin{aligned}
    \mathcal{H}_{T}[U(x,y), M_t] = & \mathscr{F}^{-1}\{\mathscr{F}\{U\left(x, y\right)\} \cdot H(v_{x}, v_{y}, D+\Delta z)\\ \times & \exp[i2\pi(v_x \Delta x+v_y \Delta y)]\} \\
    \mathcal{H}_{T}^{-1}[U(x,y), M_t] = & \mathscr{F}^{-1}\{\mathscr{F}\{U\left(x, y\right)\} \cdot H(v_{x}, v_{y}, D+\Delta z)\\ \times & \exp[i2\pi(-v_x \Delta x-v_y \Delta y)]\} \\
\end{aligned}
\end{equation}
Based on \eqref{eq_Losround}, the channel model of RBC system with translating receivers reads
\begin{equation}\label{eq_Losroundtrans}
\begin{aligned}
\mathbf{H}_{\rm{los, t}}=\mathcal{T}_{T}^{-1}\mathcal{H}_{T}^{-1}\mathcal{T}_{R}\mathcal{H}_{T}\mathcal{T}_{T}.
\end{aligned}
\end{equation}

Moreover, the movement of the receiver also includes the rotation of the output reflector around the $x$, $y$ and $z$ axes as shown in Fig.~\ref{fig:losrao}. Based on \eqref{eq_fieldpro}, the wave vector of density distribution as
\begin{equation}\label{eq_wavevector}
\varpi = 2\pi[v_x,v_y,\omega(v_x,v_y)]
\end{equation}
with $\omega(v_x,v_y) = \left(\lambda^{-2}-v_x^2-v_y^2\right)^{1 / 2}$. Defining the rotation matrix as $\mathcal{M}=\mathbf{R_{m_1}}(\theta_1)\mathbf{R_{m_2}}(\theta_2) \cdots \mathbf{R_{m_n}}(\theta_n)$, where $\mathbf{R_{m_i}}(\theta_i)$ represents the rotation matrix around $x$, $y$, or $z$-axis, $[m_1,m_2, \cdots ,m_n]\in \{x,y,z\}$, and $\theta_i$ is the rotation angle. The relationship between the wave vectors before and after rotation can be given by $\varpi = \mathcal{M}^{-1} \hat{\varpi}$. If the rotation matrix is $\begin{aligned}
        \mathcal{M}^{-1}=\left[\begin{array}{lll}
    a_1 & a_2 & a_3 \\
    a_4 & a_5 & a_6 \\
    a_7 & a_8 & a_9
    \end{array}\right],
    \end{aligned}$
the frequency-domain coordinates $(v_x,v_y)$ before rotation can be calculated as
\begin{equation}\label{eq:rotation}
\begin{aligned}
  \begin{cases}
       v_x=\mathcal{M}_{x}(\hat{v_x}, \hat{v_y})=a_1 \hat{v_x}+a_2 \hat{v_y}+a_3 \hat{w}(\hat{v_x}, \hat{v_y}) \\
    v_y=\mathcal{M}_{y}(\hat{v_x}, \hat{v_y})=a_2 \hat{v_x}+a_5 \hat{v_y}+a_6\hat{w}(\hat{v_x}, \hat{v_y})
  \end{cases}
\end{aligned}.
\end{equation}
Then, the density distribution on the rotated receiver is:
\begin{equation}\label{eq:rorationfun}
\begin{aligned}
    \mathcal{H}_{O}[U(x,y),\mathcal{M}] = &  \mathscr{F}^{-1}\{\hat{G}(\hat{v_x},\hat{v_y})|J(\hat{v_x},\hat{v_y})|\} \\
    = &\mathscr{F}^{-1}\{G(\mathcal{M}_{x}(\hat{v_x}, \hat{v_y}),\mathcal{M}_{y}(\hat{v_x}, \hat{v_y}))\\
    &\; |J(\hat{v_x},\hat{v_y})|\}
\end{aligned}
\end{equation}
where $G(x,y)$ is the angular spectrum of density distribution $U(x,y)$, and $G(x,y)=\mathcal{F}\{U(x,y)\}\cdot H(v_{x}, v_{y}, D)$. $J(\hat{v_{x}}, \hat{v_{y}})$ represents a Jacobian function for compensating the non-linearity introduced by rotational transformation,
\begin{equation}\label{eq:Jacobian}
    \begin{aligned}
    J(\hat{v_{x}}, \hat{v_{y}}) =& \left(a_2 a_6-a_3 a_5\right) \frac{\hat{v_{x}}}{\hat{w}(\hat{v_{x}}, \hat{v_{y}})} +\left(a_3 a_4-a_1 a_6\right) \\ & \frac{\hat{v_{y}}}{\hat{w}(\hat{v_{x}}, \hat{v_{y}})}+\left(a_1 a_5-a_2 a_4\right).
    \end{aligned}
    \end{equation}
For the resonant beam propagating in the reverse direction, i.e., reflected from the receiver back to the transmitter, a similar counter-rotation process as described above should be performed, where the rotation angle takes on the negative value of $\theta_i$ in $\mathcal{M}$.

Hence, the channel model for resonant beam propagation a round-trip with rotating receivers can be expressed as
\begin{equation}\label{eq_Losroundrota}
\begin{aligned}
\mathbf{H}_{\rm{los, r}}=\mathcal{T}_{T}^{-1}\mathcal{H}_{O}^{-1}\mathcal{T}_{R}\mathcal{H}_{O}\mathcal{T}_{T}.
\end{aligned}
\end{equation}
Here, the direct LoS channel model without and with misalignment has been depicted in \eqref{eq_Losround}, \eqref{eq_Losroundtrans} and \eqref{eq_Losroundrota}.

\subsection{IRS-assisted Channel Model}

\begin{figure}[t]
\centering
\includegraphics[scale=0.35]{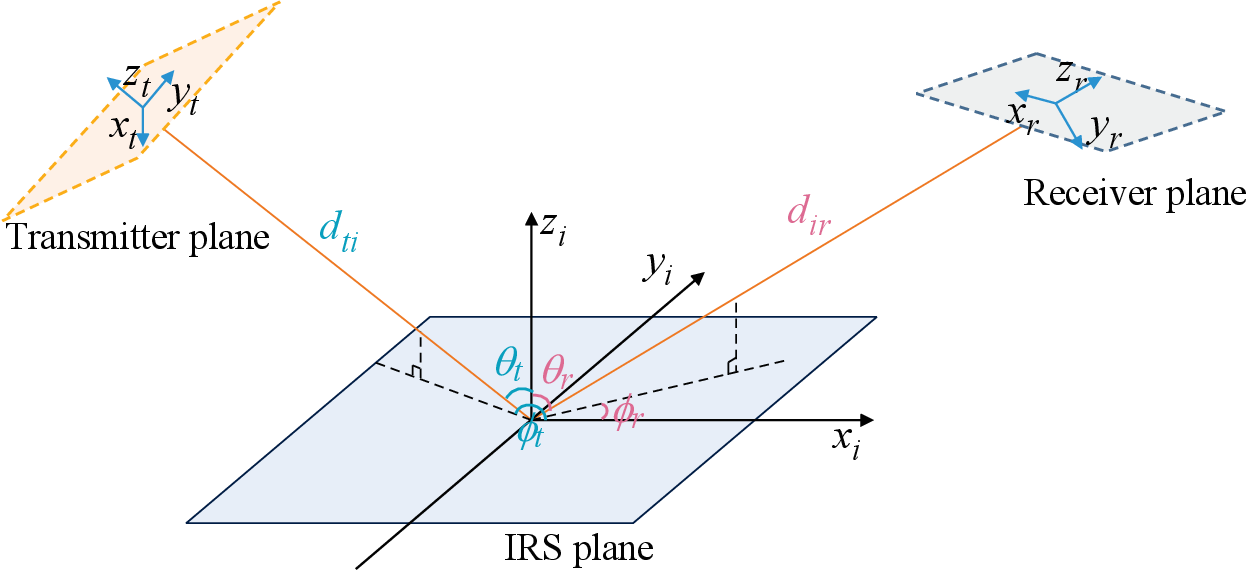}
\centering
\caption{The example coordinate system of IRS-assisted RBC system.}
\label{Fig:Car}
\vspace{-10pt}
\end{figure}

Unlike direct LoS propagation, the resonant beam passes through the IRS in free space propagation. Therefore, based on \eqref{eq_space}, the transfer function for Tx-IRS and IRS-Rx can be depicted as
\begin{equation}\label{eq_TIR}
\begin{aligned}
    \mathcal{H}_{D_{ti}}[U(x,y), D_{ti}]=\mathscr{F}^{-1}\{\mathscr{F}\{U\left(x, y\right)\} \cdot H(v_x, v_y, D_{ti})\} \\
     \mathcal{H}_{D_{ir}}[U(x,y), D_{ir}]=\mathscr{F}^{-1}\{\mathscr{F}\{U\left(x, y\right)\} \cdot H(v_x, v_y, D_{ir})\}
\end{aligned},
\end{equation}
where $D_{\rm{ti}}$ and $D_{\rm{ir}}$ are the distances of Tx-IRS and IRS-Rx, $U(x,y)$ represents the incident field distribution. Importantly, the impact of the IRS on resonant beam propagation is primarily expressed through changes in the propagation phase. Fig.~\ref{Fig:Car} introduces a sample coordinate system for the IRS-assisted RBC system, wherein the resonant beam is incident along the $z$-axis of the Tx's coordinate system onto the IRS plane, and then reflected towards the Rx. The IRS lies in the $xoy$ plane of its coordinate system, with its center at point $(x_i, y_i, 0)$, and the $z$-axis is perpendicular to the $xoy$ plane. Let $\Phi_i=(\theta_i, \phi_i)$ and $\Phi_r=(\theta_r, \phi_r)$ denote the angles of incidence and reflection at the IRS, respectively, where $\theta_k$ ($k \in {i, r}$) is the elevation angle (i.e., the angle between the beam path and the $z$-axis), and $\phi_k$ ($k \in {i, r}$) is the azimuth angle (i.e., the angle between the projection of the beam path on the $xoy$ plane and the $x$-axis). 

To ensure the consistency of coordinate systems, we rotate and transform the coordinate system of Tx. Assume that the coordinates of the emission point in the input reflector are $(x,y,z)$, and the conversion matrix is $^t_i\mathbf{R}$. The rotated coordinates in the coordinate system of the IRS plane $(x',y',z')$ are
\begin{equation}
\begin{aligned}
(x',y',z')^T &= ^t_i\mathbf{R}(x,y,z)^T \\
 &= \left[\begin{array}{ccc}
    {a_x} & {a_y} & {a_z} \\
    {b_x} & {b_y} & {b_z} \\
    {c_x} & {c_y} & {c_z}
    \end{array}\right] \left[\begin{array}{c}
    {x} \\
    {y} \\
    {z}
    \end{array}\right],
\end{aligned}
\end{equation}
where the conversion matrix can be calculated by $\mathbf{R} = \mathbf{R_x}\mathbf{R_y}\mathbf{R_z}$, $\mathbf{R_x}$ with $\mathbf{R_y}$ and $\mathbf{R_z}$ representing the counter-clockwise rotation matrix around $x$, $y$ and $z$ axes.


For the incident beam with angle $\Phi_i=(\theta_i, \phi_i)$, we transform the Tx coordinate to the Cartesian coordinate as
\begin{equation}\label{eq_corrao}
 (x_t',y_t',z_t')^T = \mathbf{R_y}(-\theta_i) \mathbf{R_z}(-\phi_i) (x_t,y_t,z_t)^T.
\end{equation}
Then, if the distances of Tx-IRS and IRS-Rx along IRS plane are known, the transmission distancess for the two stages can also be calculated as
\begin{subequations}\label{eq_tidis2}
\begin{align}
     &d_{\mathrm{ti}} = \frac{D_{x}^i}{cos(\pi-\phi_i)}\frac{1}{cos(\frac{\pi}{2}-\theta_i)}= \frac{D_{x}^i}{-cos(\phi_i)sin(\theta_i)}, \label{eq_tidis2a}\\
     &d_{\mathrm{ir}} = \frac{D_{x}^r}{cos(\phi_r)}\frac{1}{cos(\frac{\pi}{2}-\theta_r)} =\frac{D_{x}^r}{cos(\phi_r)sin(\theta_r)}, \label{eq_tidis2b}
\end{align}
\end{subequations}
where $D_{x}^i$ and $D_{x}^r$ represent the distances of Tx-IRS and IRS-Rx along the $x_i$-axis in the IRS plane.

Next, the density distribution of the resonant beam will be altered by IRS to achieve NLOS propagation,
\begin{equation}\label{eq_irsphase1}
 \mathcal{P}_{R}[U(x,y), R]=U(x,y) \cdot e^{j\Gamma(x,y)},
\end{equation}
where $(x,y)$ is the coordinate of an IRS element, $\Gamma(x,y)$ denotes the phase shift introduced by the IRS, and $\Gamma(x,y) = \frac{2\pi}{\lambda}(\Gamma_x x + \Gamma_y y)$ with a linear phase-shift profile. Given the incident angle $\Phi_i = (\theta_i,\Phi_i)$ and reflection angle $\Phi_r = = (\theta_r,\Phi_r)$, the phase-shift profile follows the well-known constant phase-shift gradient design~\cite{ajam2021channel},
\begin{subequations}\label{eq_irsphase3}
\begin{align}
      &\Gamma_x = cos(\frac{\pi}{2}-\theta_i)cos(\Phi_i)+cos(\frac{\pi}{2}-\theta_r)cos(\Phi_r), \label{eq_irsphase3a}\\
     &\Gamma_y = cos(\frac{\pi}{2}-\theta_i)sin(\Phi_i)+cos(\frac{\pi}{2}-\theta_r)sin(\Phi_r). \label{eq_irsphase3b}
\end{align}
\end{subequations}


Thus, the IRS-assisted channel function for resonant beam propagation over a round trip can be expressed as
\begin{equation}\label{eq_IRSLosround}
\begin{aligned}
\mathbf{H}_{\rm{irs}}=\mathcal{T}_{T}^{-1}\mathcal{H}_{D_{ti}}^{-1}\mathcal{P}_{R}^{-1}\mathcal{H}_{D_{ir}}^{-1}\mathcal{T}_{R} \mathcal{H}_{D_{ir}}\mathcal{P}_{R}\mathcal{H}_{D_{ti}}\mathcal{T}_{T},
\end{aligned}
\end{equation}
where $\mathcal{P}_{R}^{-1}$ represents the reverse propagation in the IRS with the opposite angle. Furthermore, the IRS-assisted channel model with receiver misalignment can be calculated based on \eqref{eq_transfun}, \eqref{eq_Losroundtrans}, \eqref{eq:rorationfun}, and \eqref{eq_Losroundrota},
\begin{subequations}\label{eq_IRSmisround}
\begin{align}
&\mathbf{H}_{\rm{irs},t}=\mathcal{T}_{T}^{-1}\mathcal{H}_{D_{ti}}^{-1}\mathcal{P}_{R}^{-1}\mathcal{H}_{T}^{-1}\mathcal{T}_{R} \mathcal{H}_{T}\mathcal{P}_{R}\mathcal{H}_{D_{ti}}\mathcal{T}_{T}, \label{eq_IRSmisrounda}\\
&\mathbf{H}_{\rm{irs},o}=\mathcal{T}_{T}^{-1}\mathcal{H}_{D_{ti}}^{-1}\mathcal{P}_{R}^{-1}\mathcal{H}_{O}^{-1}\mathcal{T}_{R} \mathcal{H}_{O}\mathcal{P}_{R}\mathcal{H}_{D_{ti}}\mathcal{T}_{T}, \label{eq_IRSmisroundb}
\end{align}
\end{subequations}
where $\mathbf{H}_{\rm{irs},t}$ and $\mathbf{H}_{\rm{irs},r}$ denote the misalignment of translation and rotation, respectively. Besides, $\mathcal{H}_{T}$ and $\mathcal{H}_{O}$ are used to replace $\mathcal{H}_{D_{ir}}$ to characterize the propagation from the IRS to Rx, with the transmission distance $D$ substituted by $D_{ir}$ in $\mathcal{H}_{T}$ and $\mathcal{H}_{O}$.


\section{Channel Capacity Characterization}
Based on the developed resonant beam channel model, this section derives the output power and channel capacity of the IRS-assisted RBC system.

\subsection{Transfer Factor}
The density distribution $U(x,y)$ on the transfer apertures during a round-trip propagation in the resonant cavity can be obtained from the channel model under different propagation restrictions in \eqref{eq_Losround}, \eqref{eq_Losroundtrans}, \eqref{eq_Losroundrota}, \eqref{eq_IRSLosround}, and \eqref{eq_IRSmisround}. However, the distribution varies during successive round-trip propagations due to the presence of diffraction loss. A steady-state distribution will emerge once the propagation time reaches a sufficient number of round-trips, where the amplitude and phase of the optical field on the same aperture remain unchanged during two consecutive transmissions, also known as the self-reproducing mode. The transfer factor can then be derived from the difference between the steady-state distributions at the $n-$th and $(n+1)-$th round-trip propagations,
\begin{equation}\label{eq:transloss}
\begin{aligned}
      \rho = \frac{\Vert U_{n+1}(x,y)\Vert_1}{\Vert U_{n}(x,y)\Vert_1}\\
\end{aligned},
\end{equation}
where $\Vert U \Vert_1$ is the $L1$ norm of $U$,  representing the sum of the amplitudes of each element in the aperture. Furthermore, the transmission loss can be calculated using the transfer factor as $\delta = 1-|\rho|^2$, where $|\cdot|^2$ denotes the beam intensity. The transfer efficiency can be further calculated as $\eta=1-\delta$.

\subsection{System Output Power}
Besides, the steady-state condition of the resonator is related to the transfer factor and gain, which can be written as
\begin{equation}\label{eq:steady}
\begin{aligned}
\mathcal{G}^2 \mathcal{R}_{i}\mathcal{R}_{o}\mathcal{\eta}_s^2\eta_{ig}\eta_{go}\eta_{og}\eta_{gi}=1,
\end{aligned}
\end{equation}
where $\eta_{ig}$, $\eta_{go}$, $\eta_{og}$, and $\eta_{gi}$ denote the backward and forward transfer efficiencies of the input reflector-gain medium and gain medium-output reflector, respectively. $\mathcal{G}$ is the gain factor representing the light intensity amplified during one transit. For homogeneously broadened lasers,
\begin{equation}\label{eq:gainfactor}
\begin{aligned}
\mathcal{G}=\exp\left[\frac{\textsl{g}_0 \ell}{1+I/I_g}\right]
\end{aligned}
\end{equation}
with $\textsl{g}_0 \ell$ being the small-signal gain, a parameter dependent on the pump power and the material of the gain medium, and $I_s$ being the saturation intensity. $\mathcal{R}_{\rm{i}}$ and $\mathcal{R}_{\rm{o}}$ denote the reflectivity of input and output reflectors. Due to the frequency multiplier SHG and information loading module embedded in M$1$, the equivalent reflectivity of the input reflector can be written as $\mathcal{R}_{\rm{i}} = R_{\rm{i}} \mathcal{T}_{\rm{S}}^2(1-\eta_{\rm{S}})^2 R_{\rm{E}}$,
where $\mathcal{T}_{\rm{S}}$ and $\eta_{\rm{S}}$ are the transmittance and transmission efficiency of SHG. $R_{\rm{E}}$ is the reflectivity of EOM. Meanwhile, $\mathcal{R}_{\rm{o}}$ can be determined by the physical reflectivity of the output reflector.

Furthermore, the output optical power can be calculated based on \eqref{eq:steady} and \eqref{eq:gainfactor} as
\begin{equation}\label{eq:outgain}
\begin{aligned}
P_{\rm{o}} & = A_{\rm{b}}\eta_{\rm{go}}I(1-\mathcal{R}_{\rm{o}}) \\
& = \frac{A_{\rm{b}} I_{\rm{s}}\left(1-\mathcal{R}_{\rm{o}}\right) \eta_{\rm{go}}}{2\mid \ln \sqrt{\mathcal{R}_{\rm{i}}\mathcal{R}_{\rm{o}} \eta_{\rm{g}}^2 \eta_{\rm{o}}} \mid}\left(\textsl{g}_0 \ell-\mid \ln \sqrt{\mathcal{R}_{\rm{i}}\mathcal{R}_{\rm{o}} \eta_{\rm{g}}^2 \eta_{\rm{o}}} \mid\right)
\end{aligned}
\end{equation}
where $A_b$ and $I$ are the cross-sectional area and intensity of the resonant beam on gain medium, and their product corresponds to the beam power on the gain medium. The small-signal gain is determined by the pump power $P_{\rm{i}}$ and the excitation efficiency $\eta_{\rm{e}}$, with the relationship $\textsl{g}0 \ell = \eta_{\rm{e}} P_{\rm{i}} / (A_{\rm{g}} I_{\rm{g}})$, where $A_{\rm{g}}$ and $I_{\rm{s}}$ denote the cross-sectional area and saturated intensity of the gain medium. $\eta_{\rm{o}}=\eta_{ig}\eta_{go}\eta_{og}\eta_{gi}$ represents the transfer efficiency of a round-trip transmission. For IRS-assisted channel, the transfer efficiency in free space from Tx to Rx is divided into Tx-IRS and IRS-Rx, $\eta_{\rm{go}}=\eta_{\rm{gr}}\eta_{irs}\eta_{\rm{ro}}$. $\eta_{\rm{gr}}$ and $\eta_{\rm{ro}}$ are the forward transfer efficiency of the two stages. $\eta_{irs}$ is the transfer efficiency of IRS.

Next, the output power of frequency-doubled beam $P_{o}^{2v}$ for communication can be derived as
\begin{subequations}\label{eq:powerdouble}
\begin{align}
&P_{t}^v = \frac{P_{\rm{o}}}{(1-\mathcal{R}_{\rm{o}})\eta_{ig}\eta_{go}}, \label{eq:powerdoublea}\\
&P_{t}^{2v} = {P_{t}^v}^2(1-R_{i}^{v})\eta_{\rm{S}}R_{E}, \label{eq:powerdoubleb}\\
&P_{o}^{2v} = P_{t}^{2v} R_{i}^{2v}\eta_{lg}\eta_{go}(1-R_{o}^{2v}), \label{eq:powerdoublec}
\end{align}
\end{subequations}
where $R_{i}^{v}$, $R_{i}^{2v}$ and $R_{o}^{2v}$ are the physical reflectivity for the frequency-fundamental beam of M$1$, frequency-doubled beam of M$1$ and M$2$. $P_{t}^v$, $P_{t}^{2v}$, and $P_{o}^{2v}$ denote the beam power of frequency-fundamental in Tx, frequency-doubled in Tx and Rx, respectively. Additionally, the transmission efficiency through the SHG $\eta_{\rm{S}}$ is determined by the physical characteristics,

\begin{equation}\label{eq:effSHG}
\begin{aligned}
\eta_{\rm{S}}=I_b\frac{8\pi^2C^2_{n}l^2_{\rm{s}}}{\varepsilon c \lambda^2 n^3},
\end{aligned}
\end{equation}
where $I_b$ is the beam intensity at the SHG crystal, expressed as $\frac{2P_{i}}{\pi\omega^2_b}$ with $\omega_b$ being the beam radius. $\varepsilon$ is the vacuum permeability, $c$ is the lightspeed, $C_{n}$, $l_{\rm{s}}$, and $n$ represent the effective nonlinear coefficient, thickness, and refractive index of the SHG crystal, respectively.
\begin{algorithm}
    \caption{Optimizing Communication Power}
    \begin{algorithmic}[1]
    \Require $z$, $l_{\rm{s}}$, $P_{i}$
    \State initialize $R_{\rm{i}}$, $R_{\rm{o}}$, $D_{x}^i$, $D_{x}^r$, $r_{\rm{M}}$, $r_{\rm{L}}$, $r_{\rm{B}}$, $M_{\rm{t}}$, $\mathcal{M}$, $d \gets 0$
    \State  $U(x,y) \gets \mathbf{H}_{\rm{los}}/\mathbf{H}_{\rm{irs}}U'(x,y)$
    \State do \eqref{eq:transloss} to get $\eta$
    \State do \eqref{eq:powerdoublea} to get $P_{\rm{t}}^{v}$
    \While {$0<d<r_{\rm{B}}$ \textbf{and} $\exists M_{\rm{t}}$ \textbf{and} $\exists \mathcal{M}$}
    \State $\eta \gets$\eqref{eq_Losround}, \eqref{eq_Losroundtrans}, \eqref{eq_Losroundrota}, \eqref{eq_IRSLosround}, \eqref{eq_IRSmisround}
    \State $P_{\rm{oc}}^{d}\gets \eqref{eq:sumoutpowera}\  |_{\gamma \gets 1}$
    \State $P_{\rm{oc}}^{i}\gets \eqref{eq:sumoutpowerb}\  |_{\gamma \gets 0}$
    \State $P_{\rm{oc}}^{m} \gets max(P_{\rm{oc}}^{d}, P_{\rm{oc}}^{i})$
    \State $P_{\rm{oc}} \gets P_{\rm{oc}}^{d} + P_{\rm{oc}}^{i}\  |_{\gamma \in [0,1]}$
    \State $\gamma \gets 1$
    \If {$P_{\rm{oc}} \leq P_{\rm{oc}}^{m}$}
    \State $\gamma \gets \gamma + \Delta\gamma$
    \State $P_{\rm{oc}}^{d}$, $P_{\rm{oc}}^{i}$ $\gets$ \eqref{eq:sumoutpower}
    \State $P_{\rm{oc}} \gets P_{\rm{oc}}^{d} + P_{\rm{oc}}^{i}$
    \EndIf
    \EndWhile
    \State \Return{$P_{\rm{oc}}$, $\gamma$}
    \label{FAFCA}
    \end{algorithmic}
    \end{algorithm}

Furthermore, the communication power is the sum of frequency-doubled beam power through direct LoS channel and IRS-assisted channel, i.e. $P_{\rm{oc}} = P_{\rm{oc}}^{\rm{d}}+P_{\rm{oc}}^{\rm{i}}$, reads
\begin{subequations}\label{eq:sumoutpower}
\begin{align}
&P_{\rm{oc}}^{\rm{d}} = \gamma P_{t}^{2v} (1-R_{i}^{2v})\eta_{\rm{ig}}\eta_{\rm{go}}(1-R_{o}^{2v}), \label{eq:sumoutpowera}\\
&P_{\rm{oc}}^{\rm{i}} = (1-\gamma) P_{t}^{2v} (1-R_{i}^{2v})\eta_{\rm{ig}}\eta_{\rm{gr}}\eta_{\rm{r}}\eta_{\rm{ro}}(1-R_{o}^{2v}), \label{eq:sumoutpowerb}
\end{align}
\end{subequations}
where $\gamma$ represents the power-splitting (PS) ratio of the two channels relating to the transfer efficiency of the direct LoS channel, and $0\leq\gamma\leq1$.

\subsection{Power Maximization and Channel Capacity}
From \eqref{eq:powerdouble} and \eqref{eq:sumoutpower}, the total power of the frequency-doubled beam $P_{t}^{2v}$ at the transmitter is determined by fixed parameters, capping the output power available for communication from both channels. To ensure that the output power remains at its peak value when transfer efficiency decreases due to object obstructions, the energy between the two channels must be optimally distributed, i.e.
\begin{subequations}\label{eq:Maximization}
\begin{align}
&\left.\max_{d, {\forall{M_t}}, {\forall{\mathcal{M}}}} P_{\rm{oc}}(\gamma)\right|_{\substack{z=z_{\text{ini}} \\ l_{\rm{s}}=l_{{\rm{s}}, \text{ini}}\\
P_{\rm{i}}=P_{{\rm{i}}, \text{ini}}}},
\label{eq:Maximizationa}\\
 s.t.\ \  &P_{\rm{oc}}\leq P_{\rm{oc}}^d(\gamma=1, \Delta d=0),  \label{eq:Maximizationb}\\
 &d \in [0, r_{\rm{B}}], \label{eq:Maximizationc}\\
     & \gamma \in [0,1], \label{eq:Maximizationd}
\end{align}
\end{subequations}
where $\forall{M_t}$, $\forall{\mathcal{M}}$ in \eqref{eq:Maximizationa} denote that the misalignment parameters, including translation and rotation, are randomized. Besides, the system parameters $z$, $l_{\rm{s}}$ and $P_{\rm{i}}$ are preset. \eqref{eq:Maximizationa} represents that the maximum power is less than the output power in the direct LoS channel when the total power is transmitted through the channel without external obstructions. \eqref{eq:Maximizationb} restricts the obstruction depth $d$ in the LoS channel to the aperture radius, which is equal to the radii of input/output reflectors. \eqref{eq:Maximizationc} states that $\gamma$ must be between 0 and 1.

Under the preset system parameters and PS ratios, the total output power $P_{\rm{oc}}$ needs to be monitored in real-time in response to external object obstruction and receiver movement. The monitored power should then be compared with the output power fully allocated to the direct LoS channel ($\gamma=1$) and the IRS-assisted channel ($\gamma=0$) under these conditions. If the total output power is less than the output power of the two independent channels, the PS ratio should be dynamically adjusted to maximize the output power. We adopt the gradient ascent method to solve the maximization power, the algorithm is depicted as Algorithm 1.

\begin{figure*}[!t]
\centering
\subfigure[Transmission distance]{\includegraphics[width=1.85in]{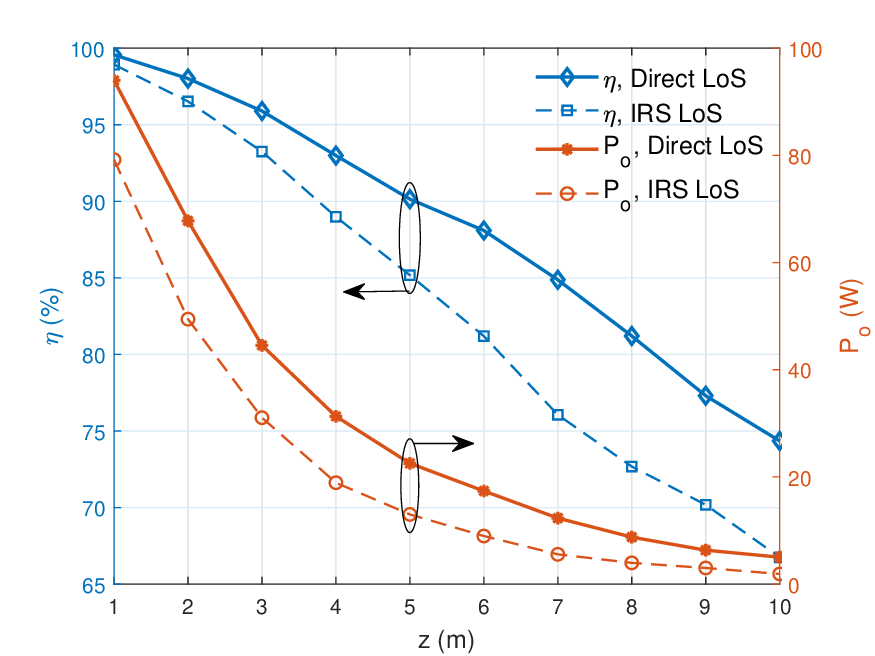}
\label{fig_Efficiency1}}
\hspace{-3mm}
\subfigure[Invasion depth]{\includegraphics[width=1.85in]{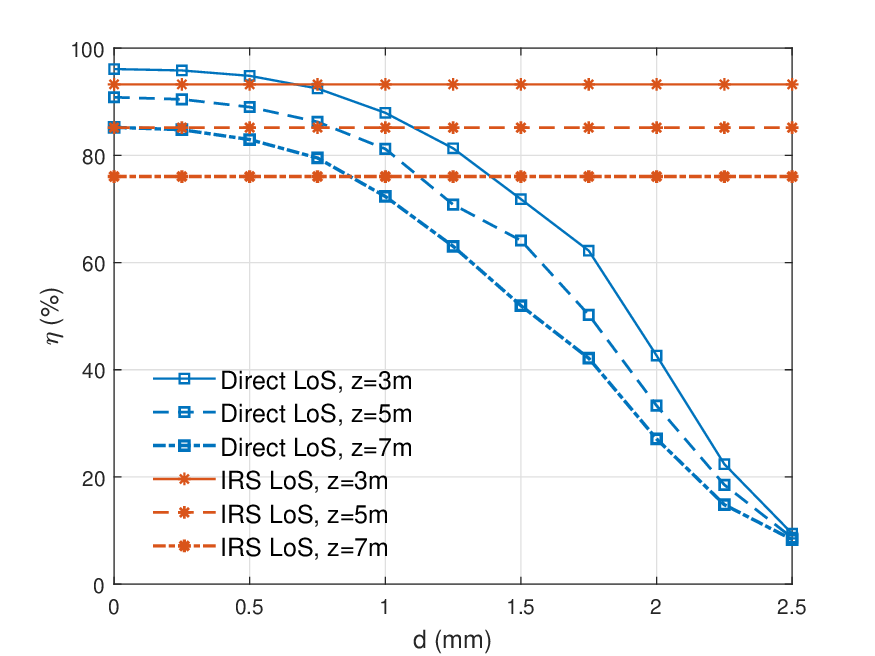}
\label{fig_Efficiency2}}
\hspace{-8mm}
\subfigure[Translation distance (z=5m)]{\includegraphics[width=1.85in]{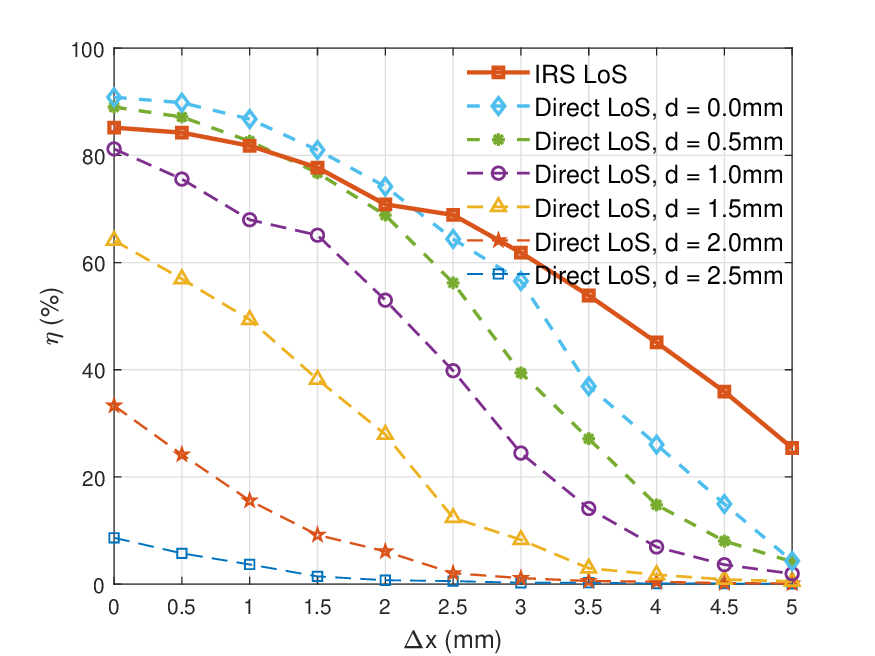}
\label{fig_Efficiency3}}
\hspace{-8mm}
\subfigure[Rotation angle (z=5m)]{\includegraphics[width=1.85in]{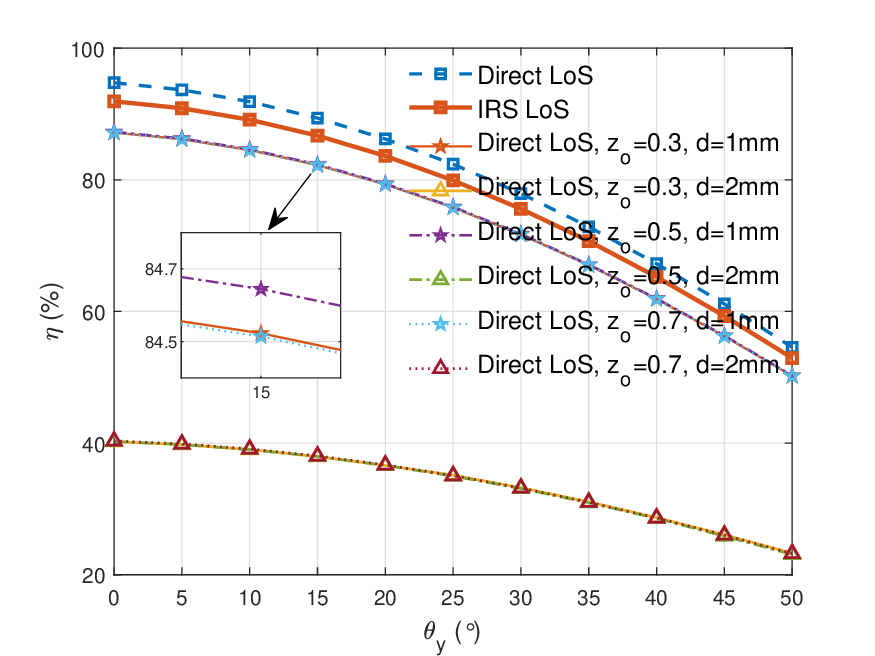}
\label{fig_Efficiency4}}
\caption{The changes of E2E efficiency $\eta$ in direct LoS and IRS-assisted channels under different parameters ($P_{\rm{in}}=200$ W).}
\label{fig_Efficiency}
\end{figure*}

Based on the optimized output power under different conditions, the signal-to-noise ratio (SNR) and channel capacity can be derived from the received communication power of two channels,
\begin{equation}\label{eq:SNR}
\begin{aligned}
 \mathrm{SNR} = \frac{(\eta_{c} P_{\rm{oc}})^2}{2\pi e \left[2 q\left(\eta_{c} P_{\mathrm{oc}}+I_{\mathrm{k}}\right) B+\frac{4 K T B}{L_{\mathrm{r}}}\right]}
\end{aligned},
\end{equation}
where $\eta_c$ represents the photo-detector responsivity in $A/W$, and $e$ is the natural constant. $[\cdot]$ denotes the total noise, expressed as $\sigma^2$, which includes background current $I_{\rm{k}}$, noise bandwidth $B$, load resistance $L_{r}$, electric charge $q$, Boltzmann's constant $K$, and temperature in Kelvin $T$.

Then, the channel capacity, represented by the spectral efficiency $S_{\rm{E}}$, can be expressed as
\begin{equation}\label{eq:capacity}
\begin{aligned}
 S_{\rm{E}} = B_{\rm{c}}\log_2\left(1+\mathrm{SNR}\right)
\end{aligned},
\end{equation}
where $B_{\rm{c}}$ denotes the channel bandwidth.

\section{Numerical Evaluation}
In this section, we evaluate the performance of RIS-assisted RBC system, including end-to-end (E2E) efficiency, output power, the frequency-doubled beam power, and channel capacity for communication under different parameters.

The structural parameters of the system are as follows: the radii of the input/output reflectors and the gain medium are $2.5$ mm, the distance between M$1$/M$2$ and L$1$/L$2$ is $5$ cm, and the saturation intensity for the gain medium with Nd:YVO$_4$ crystal is $1260$ $\rm{W/cm^2}$~\cite{hodgson2005laser}. We adopt the LiNbO$_3$ crystal as the SHG medium for frequency doubling, which has a high nonlinear coefficient of $4.7$ pm/V and a walk-off angle of $0^{\circ}$ at $1064$ nm; the refractive index of LiNbO$_3$ SHG is $2.23$~\cite{koechner2013solid}. For channel capacity calculation, the values of the parameters in \eqref{eq:SNR} are: $\eta{\rm{c}} = 0.6$ A/W, $I{\rm{c}} = 5100$ $\mu$A, $B = 811.7$ MHz, $L{\rm{r}}=5100$ K$\Omega$, and $K=300$ K. Finally, to maintain the self-reproducing model between transceivers, we assume the reflectivity of M$1$ for the fundamental frequency beam is $5\%$, allowing the remaining $95\%$ of the transmitted beam to be frequency-doubled. The wavelength of the fundamental frequency is $1064$ nm, resulting in a frequency-doubled wavelength of $532$ nm.
\subsection{End-to-End Efficiency}
In the IRS-assisted RBC system, the primary factor affecting communication performance is channel transmission efficiency. Thus, we first analyze the variations in E2E (Tx-Rx) efficiency $\eta$ ($=\sqrt{\eta_{\rm{o}}}$).

Figure~\ref{fig_Efficiency1} illustrates the impact of spatial distance $z$ on $\eta$ and output power $P_{o}$ in both the direct LoS channel and the IRS-assisted channel. It is observed that both efficiency and output power decrease with increasing transmission distance, with $P_{o}$ decreasing from its peak value to nearly $0$ W. Moreover, compared to the direct LoS channel, $\eta$ and $P_{\rm{o}}$ in the IRS-assisted channel are notably lower due to the longer path of the resonant beam. For instance, at a distance of $z=5$ m, $\eta$ is $90.14\%$ with $P_{\rm{o}}=22.54$ W in the direct LoS channel, whereas in the IRS-assisted LoS channel, $\eta$ is $85.18\%$ with $P_{\rm{o}}=13.01$ W. Furthermore, in the direct LoS channel, obstruction by external objects decreases transfer efficiency, as illustrated in Fig.~\ref{fig_Efficiency2}, due to the narrowing effective transfer aperture with increasing obstruction depth $d$. The E2E efficiency in the direct LoS channel noticeably decreases as $d$ increases from $0$ to $2.5$ mm, regardless of transmission distance. Conversely, active modulation of IRS enables the IRS-assisted channel to circumvent object occlusion effects, maintaining constant $\eta$ regardless of $d$.

\begin{figure*}[!t]
\centering
\subfigure[Invasion depth]{\includegraphics[width=2.3in]{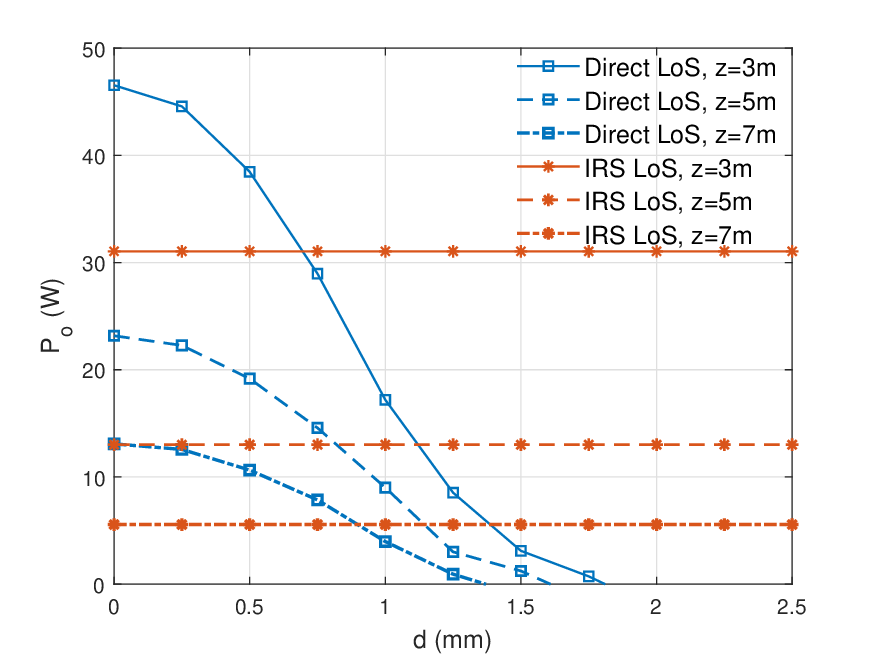}
\label{fig_power1}}
\hfil
\subfigure[Translation distance]{\includegraphics[width=2.3in]{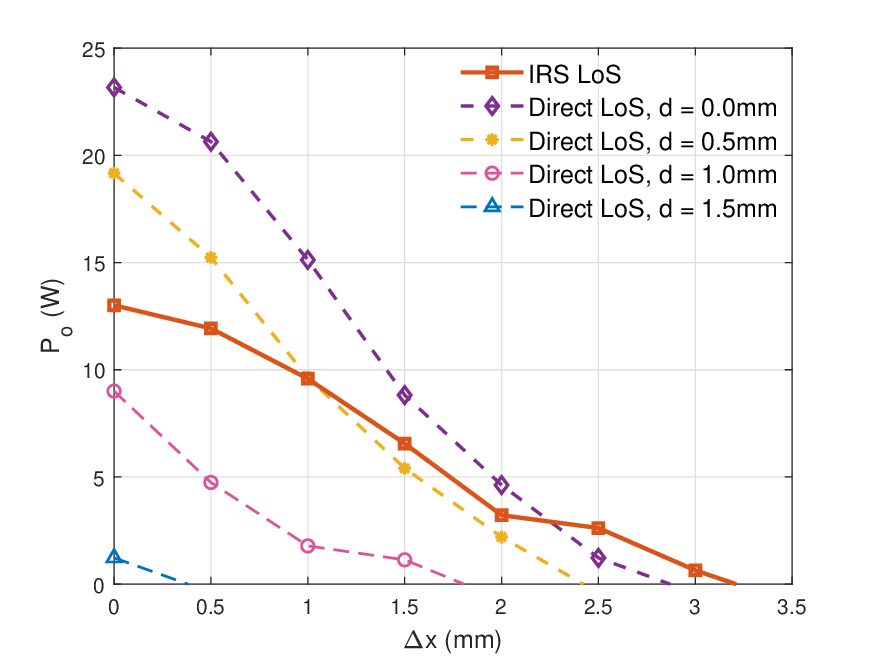}
\label{fig_power2}}
\hfil
\subfigure[Ratation angle ($d=1$ mm)]{\includegraphics[width=2.3in]{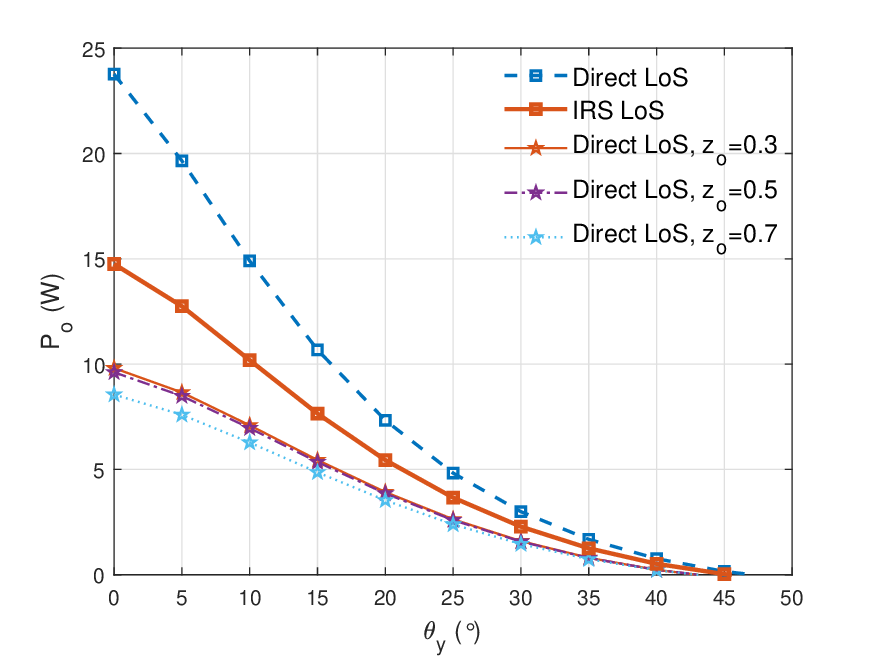}
\label{fig_power3}}
\caption{The changes of output power $P_{\rm{o}}$ in direct LoS and IRS-assisted channels with different parameter ($z=5$ m, $P_{\rm{in}}=200$ W).}
\label{fig_power}
\end{figure*}

\begin{figure*}[!t]
\centering
\subfigure[Input power ($l_s=2$ mm)]{\includegraphics[width=1.9in]{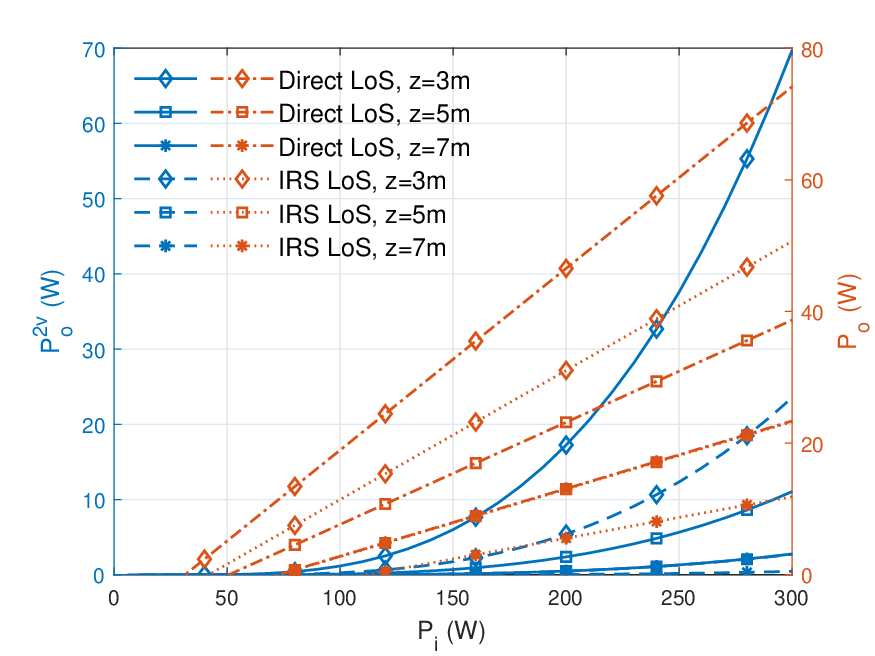}
\label{fig_doublepower1}}
\hspace{-8mm}
\subfigure[SHG thickness]{\includegraphics[width=1.9in]{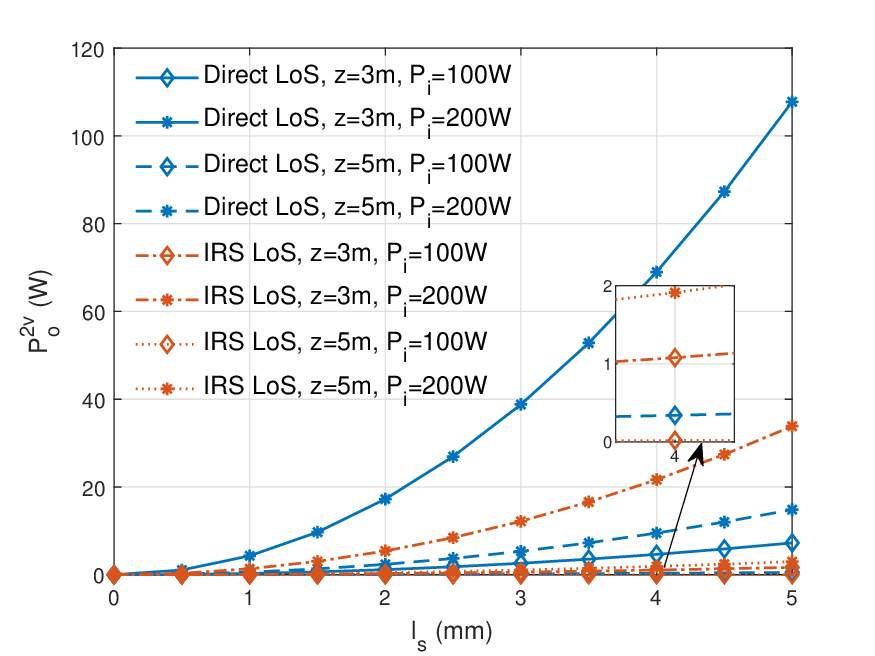}
\label{fig_doublepower2}}
\hspace{-8mm}
\subfigure[Translation distance]{\includegraphics[width=1.9in]{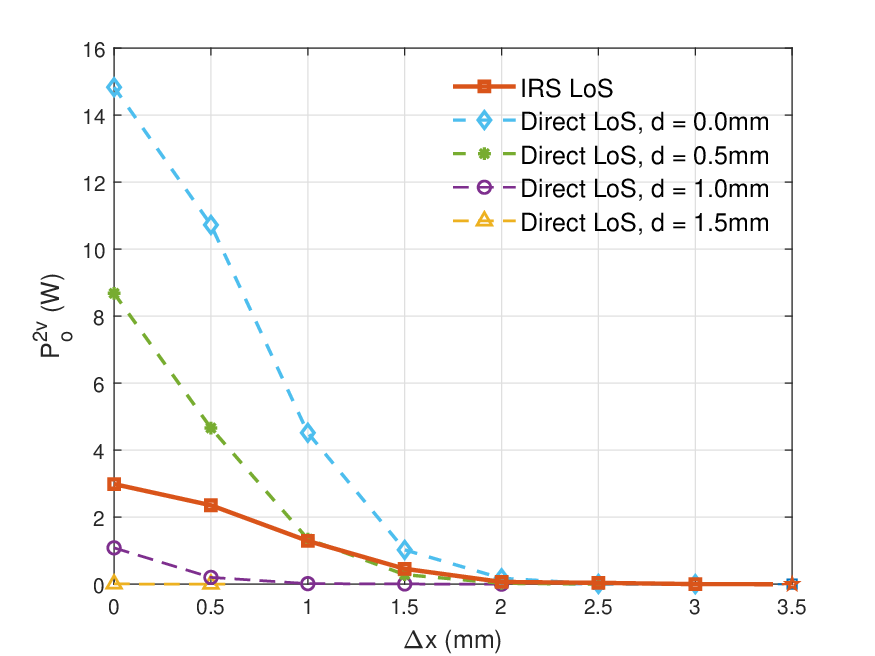}
\label{fig_doublepower3}}
\hspace{-8mm}
\subfigure[Rotation angle]{\includegraphics[width=1.9in]{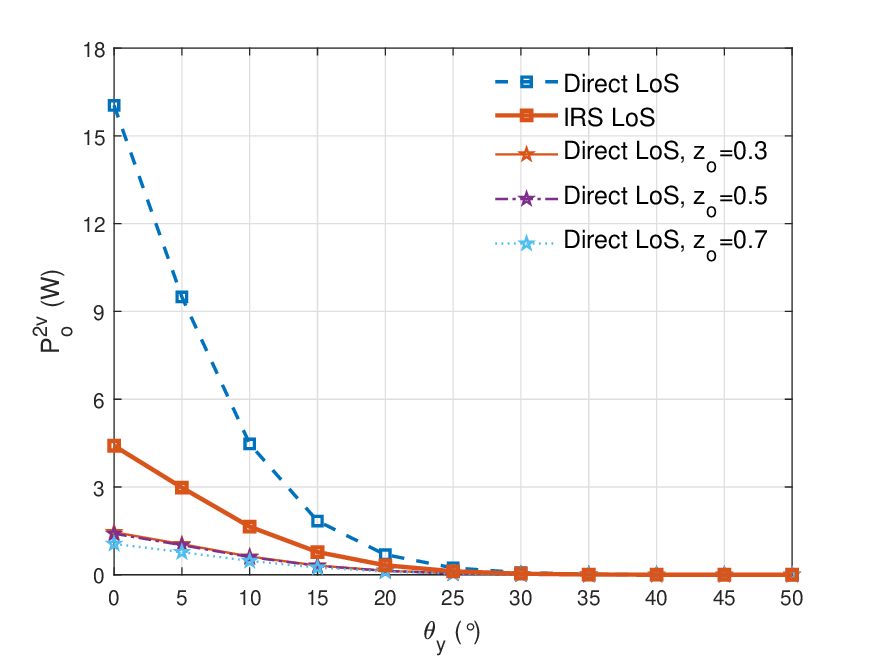}
\label{fig_doublepower4}}
\caption{The changes of the output frequency-doubled beam power $P_{\rm{o}}^{2v}$ in direct LoS and IRS-assisted channels with different parameters.}
\label{fig_doublepower}
\vspace{-10pt}
\end{figure*}

Moreover, the effects of receiver misalignment on efficiency are depicted in Figs.~\ref{fig_Efficiency3} and~\ref{fig_Efficiency4}. Under different obstruction depths $d$, E2E efficiency in both channels gradually decreases as the translation distance $\Delta x$ increases. For the direct LoS channel, deeper obstructions lead to lower transmission efficiency; for example, at $d=1$ mm, $\eta$ is significantly lower than at $d=2$ mm. Specifically, at $\Delta x=2$ mm, $\eta$ is $52.92\%$ at $d=1$ mm and $6.08\%$ at $d=2$ mm. Importantly, when the direct LoS channel is free of obstacles or has minimal interference ($d\leq0.5$ mm), the IRS-assisted channel performs worse than the direct LoS channel at short translation distances ($\Delta\leq2$ mm). Conversely, if the translation distance exceeds $2.5$ mm, the efficiency of the IRS-assisted channel surpasses that of the direct LoS channel because IRS phase adjustments can compensate for losses at greater distances. Moreover, the superiority of the IRS-assisted channel becomes more pronounced with increasing translation distance; for example, the efficiency difference between IRS-assisted and direct LoS channels without obstruction is $5.29\%$ at $\Delta x=3$ mm and increases to $19.06\%$ at $\Delta x=4$ mm. Finally, regarding misalignment in rotation, $\eta$ decreases gradually with increasing rotation angle due to the diminishing effective aperture. At the same obstruction depth, $\eta$ is significantly lower at $d=2$ mm than at $d=1$ mm for the same rotation angle, indicating that greater obstruction depth results in lower efficiency. Meanwhile, the invasion position along the optical axis $z_{\rm{o}}$ (defined as the distance between Tx and external object divided by $z$) has a relatively minor impact on efficiency; curves of $\eta$ at $z_{\rm{o}}=0.3$, $0.5$, and $0.7$ with $d=1$ mm and $2$ mm are nearly identical. Moreover, under the same rotation angle, $\eta$ in the direct LoS channel without obstruction ($d=0$ mm) is higher than in the IRS-assisted channel due to the shorter transmission distance. For example, at $\theta_y=10^{\circ}$ and $20^{\circ}$, $\eta$ is $91.88\%$ and $86.21\%$ in the direct LoS channel with $d=0$ mm, compared to $89.13\%$ and $83.63\%$ in the IRS-assisted channel.

\subsection{Power for Communication}
To analyze the communication power with the frequency-doubled beam, we first evaluate the power in the fundamental frequency under different parameters, as depicted in Fig.\ref{fig_power}. In Fig.\ref{fig_power1}, the output power $P_{\rm{o}}$ in the direct LoS channel decreases with increasing obstruction depth $d$, whereas in the IRS-assisted channel, it remains constant regardless of transmission distance. Similar to the trend in $\eta$ shown in Fig.~\ref{fig_Efficiency2}, longer distances result in lower output power. Specifically, $P_{\rm{o}}$ for both channels at $z=3$ m surpasses that at $z=5$ m, which in turn exceeds $P_{\rm{o}}$ at $z=7$ m. Furthermore, when $d$ reaches a certain value in the direct LoS channel, the output power drops to zero, as the system gain at Tx cannot compensate for the transmission losses. The invasion depth at which $P_{\rm{o}}$ reaches zero is inversely related to transmission distance, i.e., $d$ decreases at $P_{\rm{o}}=0$ as $z$ increases. For example, $d\approx1.3$ mm at $z=7$ m, $d\approx1.6$ mm at $z=5$ m, and $d\approx1.8$ mm at $z=3$ m.
Figure~\ref{fig_power2} shows the changes in output power with translation misalignment. As $\Delta x$ increases, the output power decreases. In the direct LoS channel, $P_{\rm{o}}$ drops with increasing depth at the same translation distance. Then, $P_{\rm{o}}$ of the direct LoS channel is superior to that of the IRS-assisted channel when $\Delta x$ and $d$ are small. It can be seen that $P_{\rm{o}}$ at $\Delta x\leq2.25$ mm and $1$ mm in direct LoS channel with $d=0$ mm and $0.5$ mm is greater than that in IRS-assisted channel. The variation of output power with rotation angle is illustrated in Fig.~\ref{fig_power3}, where $P_{\rm{o}}$ diminishes as $\theta_y$ rises from $0$ to $50^{\circ}$. The output power in the direct LoS channel without invasion is significantly higher than in the IRS-assisted channel, regardless of the rotation angle. For instance, $P_{\rm{o}}$ is $19.65$ W and $10.68$ W at $\theta_y=10^{\circ}$ and $20^{\circ}$ in the direct LoS channel, compared to $12.75$ W and $7.65$ W in the IRS-assisted channel.

\begin{figure*}[!t]
\centering
\subfigure[SHG thickness]{\includegraphics[width=1.9in]{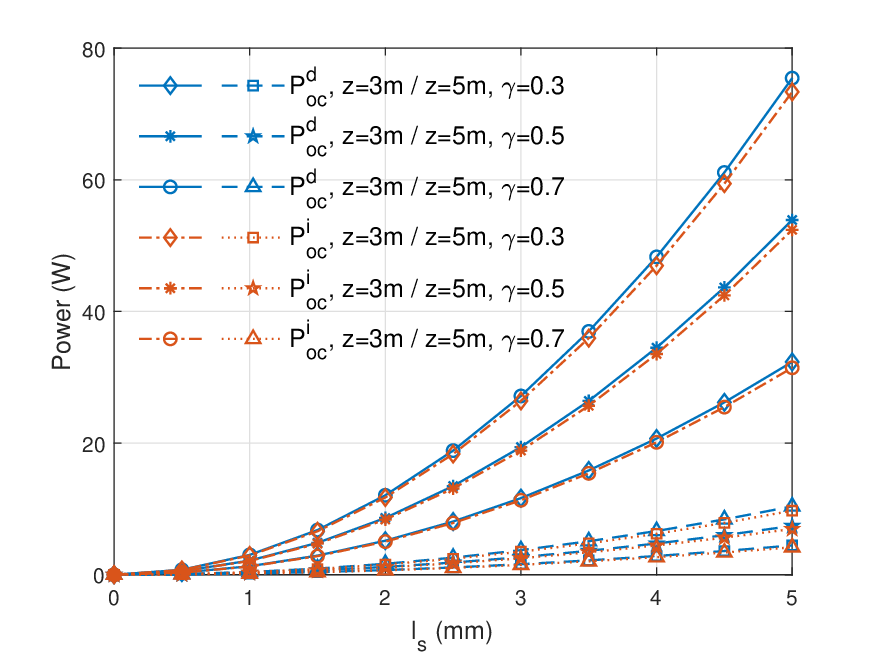}
\label{fig_compower1}}
\hspace{-8mm}
\subfigure[Invasion depth]{\includegraphics[width=1.9in]{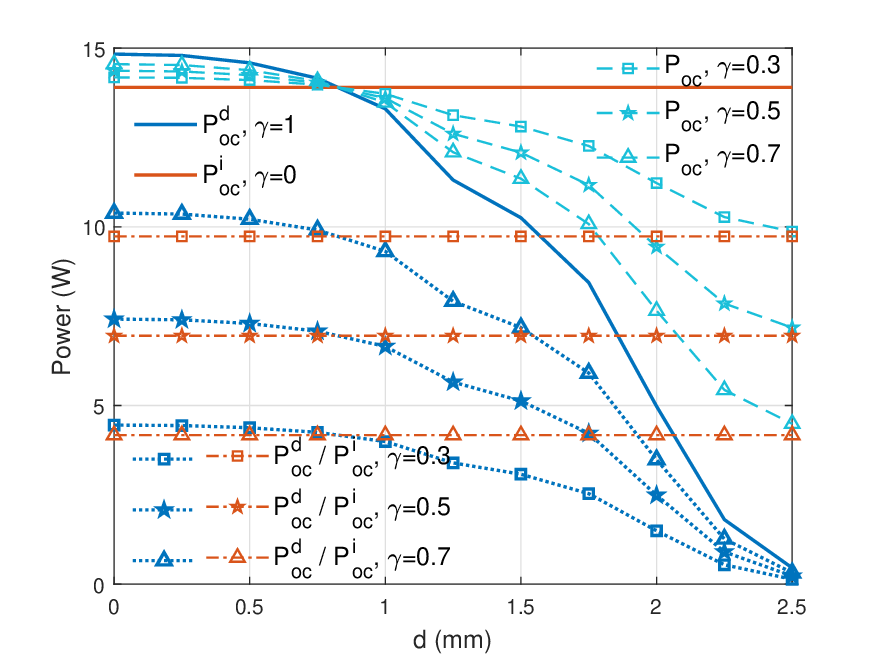}
\label{fig_compower2}}
\hspace{-8mm}
\subfigure[PS ratio with translation]{\includegraphics[width=1.9in]{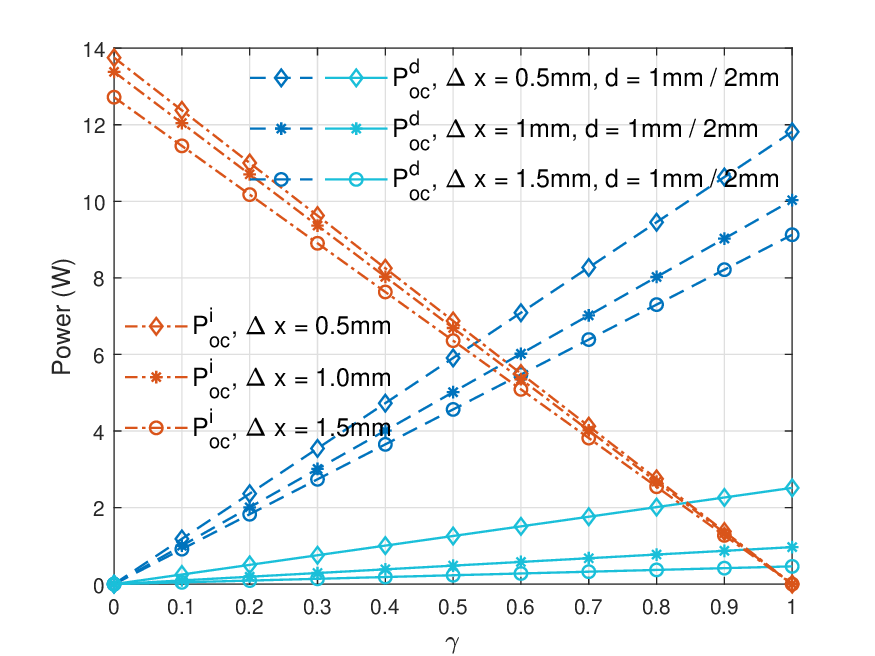}
\label{fig_compower3}}
\hspace{-8mm}
\subfigure[PS ratio with rotation]{\includegraphics[width=1.9in]{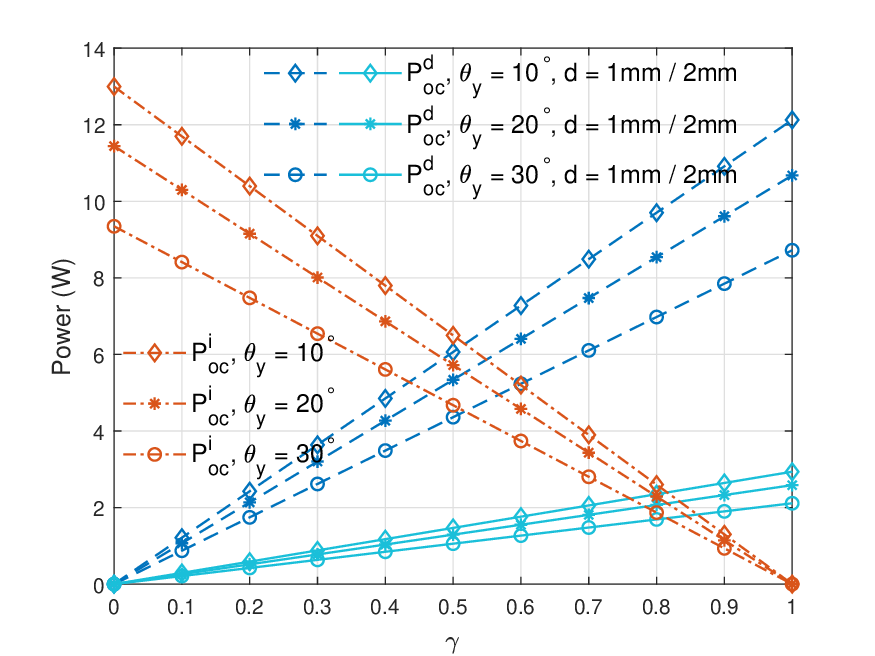}
\label{fig_compower4}}
\caption{The changes of the communication power in two channels $P_{\rm{oc}}^d$ and $P_{\rm{oc}}^i$ with different PS ratios ($P_{i}=200$ W, $z=5$ m in b, c, d).}
\label{fig_compower}
\end{figure*}

\begin{figure*}[!t]
\centering
\subfigure[Invasion depth]{\includegraphics[width=2.3in]{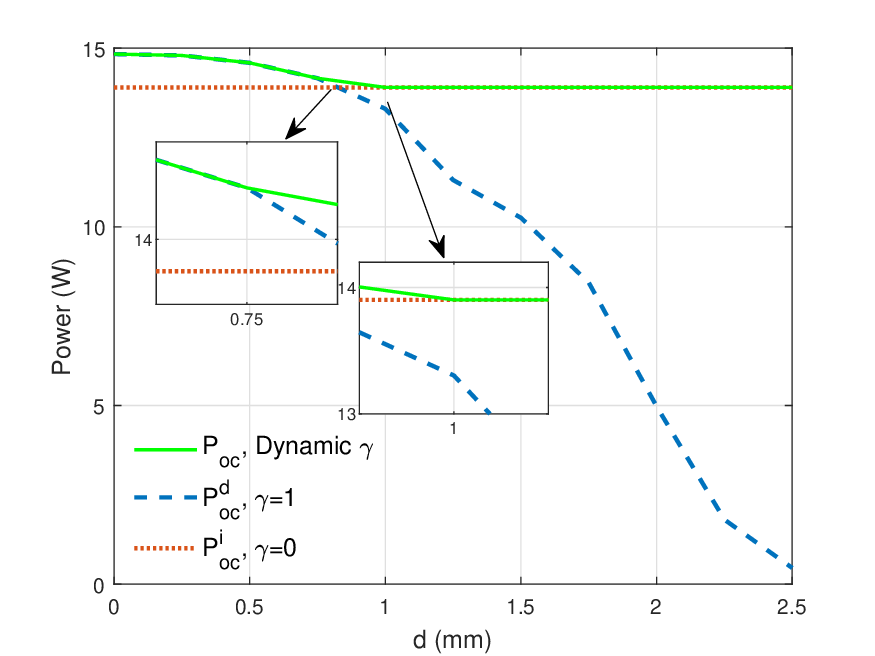}
\label{fig_dynamicpower1}}
\hfil
\subfigure[Translation distance]{\includegraphics[width=2.3in]{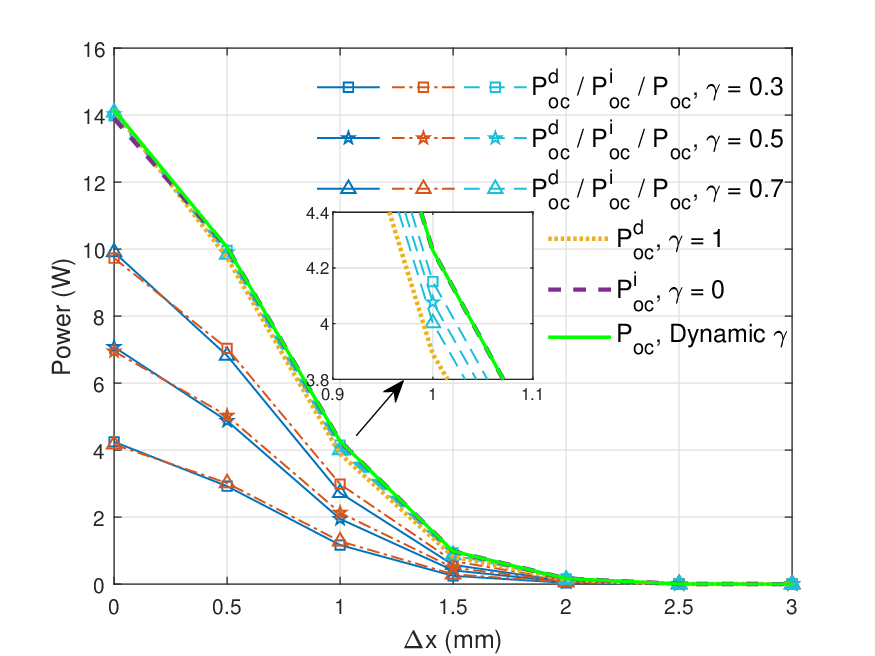}
\label{fig_dynamicpower2}}
\hfil
\subfigure[Rotation angle]{\includegraphics[width=2.3in]{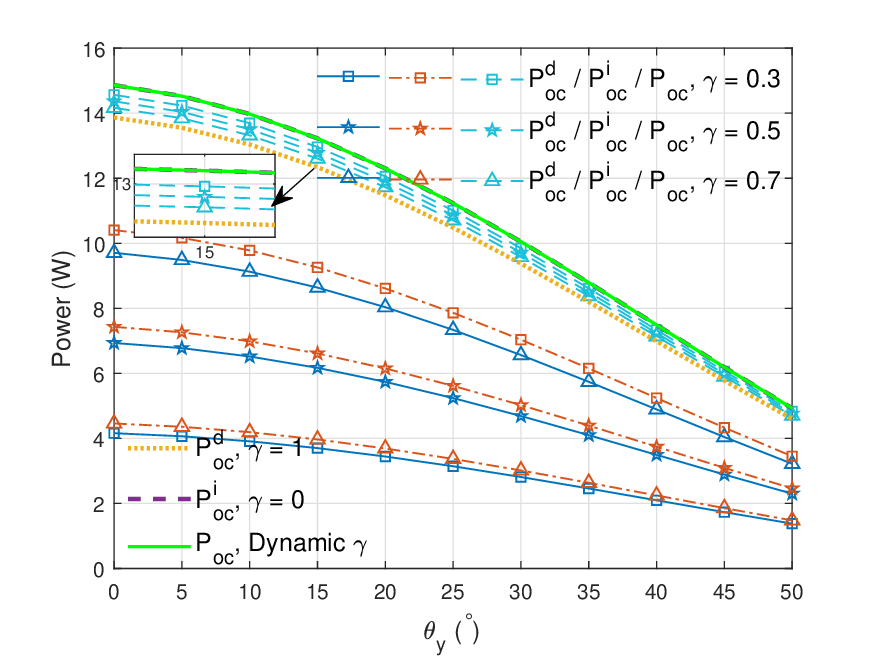}
\label{fig_dynamicpower3}}
\caption{The impact of dynamic power-splitting ratio on the communications under different parameters ($z=5$ m, $P_{\rm{i}}=200$ W).}
\label{fig_dynamicpower}
\end{figure*}

Then, the output power for frequency-doubled beam $P_{\rm{o}}^{2v}$ can be derived from \eqref{eq:powerdouble} based on transfer efficiency and output power for frequency-fundamental beam. Figure~\ref{fig_doublepower}  illustrates the variations in frequency-doubled power under different input power $P_{\rm{i}}$, SHG thickness $l_{\rm{s}}$, and misalignment parameters $\Delta x$ and $\theta_y$, assuming all power is allocated to one channel.

Firstly, the output power for frequency-fundamental $P_{\rm{o}}$ and frequency-doubled $P_{\rm{o}}^{2v}$ gradually enhance as the input power $P_{\rm{i}}$ rises from $0$ to $300$ W as illustrated in Fig.~\ref{fig_doublepower1}, and the output power remains essentially $0$ W when $P_{i}$ is less than $50$ W since the pumping power threshold is not yet satisfied. $P_{\rm{o}}^{2v}$ and $P_{\rm{o}}$ in the direct LoS channel are greater than those in the IRS-assisted channel under the same input power and distance, due to lower transmission losses. For example, $P_{\rm{o}}^{2v}$ is $1.16$ W, $17.24$ W, $69.74$ W in direct LoS channel with $P_{\rm{i}}=100$ W, $200$ W, $300$ W and $z=3$ m, compared to $0.26$ W, $5.41$ W, and $23.54$ W in the IRS-assisted channel. As shown in Fig.~\ref{fig_doublepower2},  the thickness of the SHG crystal is another significant factor affecting the frequency-doubled power. The power increases with the SHG thickness regardless of transmission distance and pump power. For the same $l_{\rm{s}}$, $P_{\rm{o}}^{2v}$ in the direct LoS channel is also greater than in the IRS channel. At $l_{\rm{s}}=2$ mm, $3$ mm, and $4$ mm, with $z=3$ m and $P_{\rm{i}}=200$ W, $P_{\rm{o}}^{2v}$ is $17.25$ W, $38.80$ W, and $68.98$ W in the direct LoS channel, compared to $5.41$ W, $12.18$ W, and $21.66$ W in the IRS channel. Afterwards, the frequency-doubled power with misalignment receiver at $l_{\rm{s}}=5$ mm, $P_{\rm{i}}=200$ W and $z=5$ m is depicted in Figs.~\ref{fig_doublepower3} and~\ref{fig_doublepower4}, where the power drops from the maximum with the increase in $\Delta x$ and $\theta_y$. Additionally, in the direct LoS channel, the power decreases with increasing obstruction depth $d$, dropping to $0$ more rapidly with greater $d$.
For example, $P_{\rm{o}}^{2v}$ is $10.72$ W, $4.66$ W, $0.2$ W at $d=0$ mm, $0.5$ mm, and $1.0$ mm under $\Delta x=0.5$ mm. When $d=0.5$ mm and $1.5$ mm, the translation distances at which $P_{\rm{o}}^{2v}$ drops to $0$ are approximately $2$ mm and $1.5$ mm, respectively. Similar to the fundamental frequency power in Fig.\ref{fig_power2}, the power in the IRS-assisted channel is lower than in the direct LoS channel if the obstruction depth and translation distance are small. Finally, $P_{\rm{o}}^{2v}$ in the direct LoS channel is greater than in the IRS-assisted channel if the rotation angle is less than a certain value, i.e., $\theta_y<30^{\circ}$. For instance, $P_{\rm{o}}^{2v}$ is $9.50$ W, $4.47$ W, and $1.84$ W at $\theta_y=5^{\circ}$, $10^{\circ}$, and $15^{\circ}$ in the direct LoS channel, compared to $2.98$ W, $1.64$ W, and $0.78$ W in the IRS channel. The invasion location $z_{\rm{o}}$ has a relatively minor impact on $P_{\rm{o}}^{2v}$, as shown in Fig.\ref{fig_doublepower4}, where $P_{\rm{o}}^{2v}$ exhibits a similar trend with rotation angle at $z_{\rm{o}}=0.3$, $0.5$, and $0.7$.

Furthermore, based on the power-splitting ratio $\gamma$, the frequency-doubled beam in the transmitter can be divided into two streams, transmitting in the direct LoS channel and the IRS-assisted channel. The relationships between the output frequency-doubled power for the two channels and the PS ratio are depicted in Fig.~\ref{fig_compower}.

Firstly, as the SHG thickness $l_{\rm{s}}$ increases, the communication power in both channels, $P_{\rm{oc}}^{\rm{d}}$ and $P_{\rm{oc}}^{\rm{i}}$, rises from zero as shown in Fig.~\ref{fig_doublepower1}. Similarly, the increase in distance results in a decrease in $P_{\rm{oc}}^{\rm{d}}$ and $P_{\rm{oc}}^{\rm{i}}$. More importantly, $P_{\rm{oc}}^{\rm{d}}$ increases while $P_{\rm{oc}}^{\rm{i}}$ decreases as $\gamma$ gets greater under the same $l_{\rm{s}}$ and $z$. In other words, $P_{\rm{oc}}^{\rm{d}}$ and $P_{\rm{oc}}^{\rm{i}}$ are inversely proportional to $\gamma$. Then, the changes of $P_{\rm{oc}}^{\rm{d}}$ and $P_{\rm{oc}}^{\rm{i}}$ with the invasion depth and PS ratio are depicted in Fig.~\ref{fig_compower2}, where $P_{\rm{oc}}^{\rm{d}}$ and $P_{\rm{oc}}^{\rm{i}}$ drop from the peak power as the obstruction deepens. It is evident that as $\gamma$ changes from $0.3$, $0.5$ to $0.7$, $P_{\rm{oc}}^{\rm{d}}$ gradually increases while $P_{\rm{oc}}^{\rm{i}}$ gradually decreases under the same $d$. The sum of the power in the two channels, $P_{\rm{oc}}$, is shown by the dashed lines. When $d$ is small, the total power $P_{\rm{oc}}$ with higher $\gamma$ is greater, while $P_{\rm{oc}}$ with smaller $\gamma$ is greater if the invasion depth deepens. That is, when $d \leq 0.75$ mm, $P_{\rm{oc}}$ at $\gamma=0.7$ is higher than that at $\gamma=0.5$ and $0.3$. Conversely, when $d > 0.75$ mm, $P_{\rm{oc}}$ at $\gamma=0.3$ is greater than that at $\gamma=0.5$ and $0.7$. This is because when $d$ is small, $P_{\rm{oc}}^{\rm{d}}$ with higher $\gamma$ is greater than $P_{\rm{oc}}^{\rm{i}}$, hence the total power with higher $\gamma$ is more prominent. Conversely, as $d$ increases, $P_{\rm{oc}}^{\rm{i}}$, which constitutes a larger proportion when $\gamma$ is small, becomes greater, hence $P_{\rm{oc}}$ with smaller $\gamma$ generates a greater value. Besides, if all transmitted power is allocated to one of the channels, i.e., $P_{\rm{oc}}^{\rm{d}}$ with $\gamma=1$ or $P_{\rm{oc}}^{\rm{i}}$ with $1-\gamma=1$, the power in the direct LoS channel or IRS channel is greater than the power partially allocated, such as $\gamma = 0.3$, $0.5$, or $0.7$. Figures~\ref{fig_compower3} and~\ref{fig_compower4} show the changes in communication power with $\gamma$ under misalignment. $P_{\rm{oc}}^{\rm{d}}$ increases while $P_{\rm{oc}}^{\rm{i}}$ decreases as $\gamma$ rises from $0$ to $1$. The increased invasion depth results in a significant decrease in $P_{\rm{oc}}^{\rm{d}}$. Simultaneously, as the translation distance $\Delta x$ and rotation angle $\theta_y$ increase, the power gradually diminishes. Moreover, there are intersection points between $P_{\rm{oc}}^{\rm{d}}$ and $P_{\rm{oc}}^{\rm{i}}$ at different values of $\gamma$, which demonstrates that $\gamma$ needs to be adjusted to maximize the communication power under different misalignment and invasion parameters.

\begin{figure}[!t]
\centering
\subfigure[SNR vs. Invasion depth.]{\includegraphics[width=3.0in]{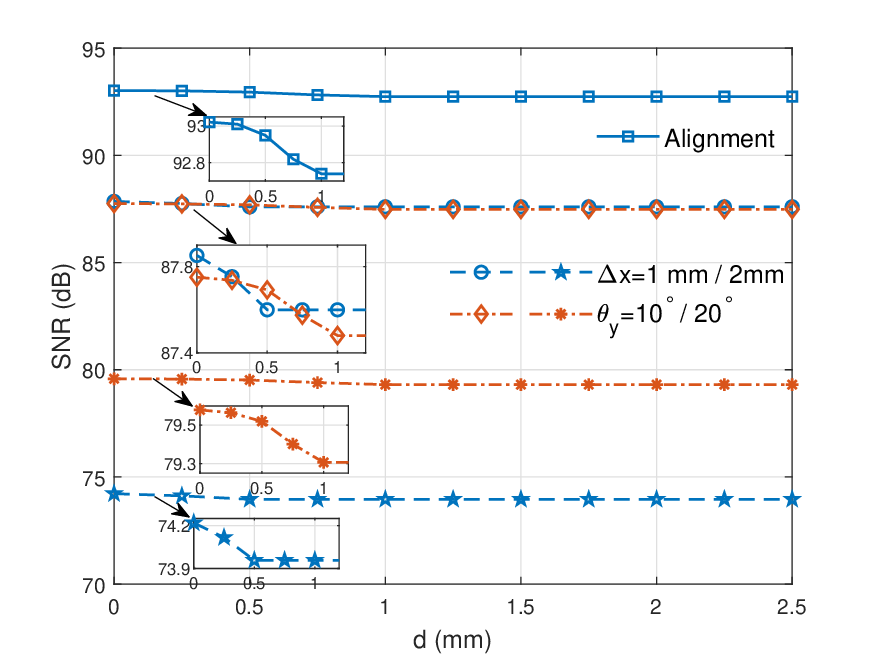}
\label{Fig_com1}}
\\ \vspace{-5pt}
\subfigure[Spectral Efficiency vs. Invasion depth.]{\includegraphics[width=3.0in]{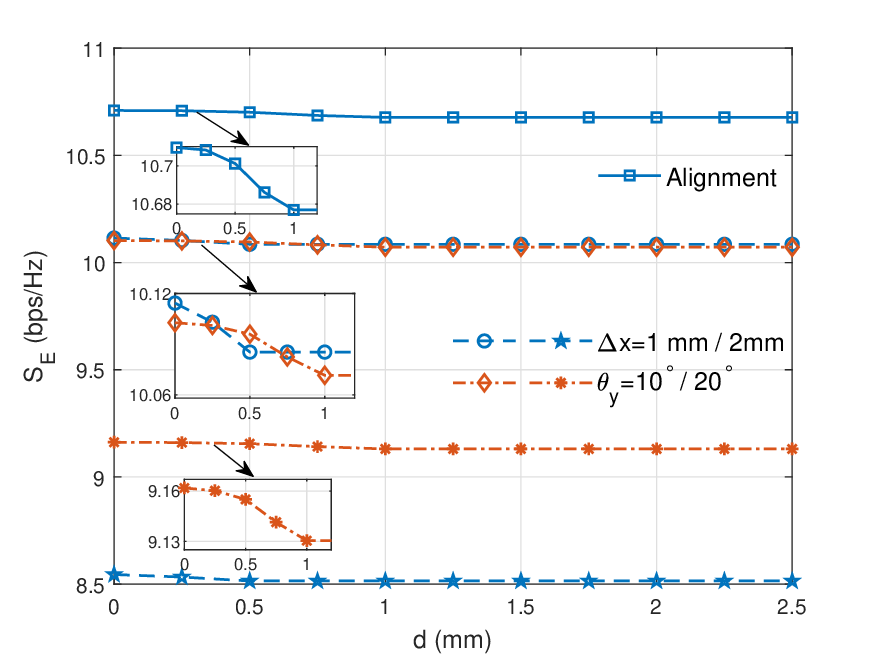}
\label{Fig_com2}}
\caption{The changes of SNR, channel capacity for communication with the optimal PS ratios.}
\label{Fig_com}
\vspace{-5pt}
\end{figure}

\subsection{Power Maximization}
To ensure optimal communication performance, the channel PS ratio between direct and IRS-assisted channels needs to be dynamically adjusted under various parameters. Next, we analyze how the maximum performance can be achieved by regulating the PS ratio.

Figure~\ref{fig_dynamicpower} depicts the impact of dynamic $\gamma$ on communication power under $z=5$ m, $l_{\rm{s}}=5$ mm, and $P_{\rm{i}}=200$ W, where the green solid line represents the maximum power after dynamic adjustment.  Based on Fig.~\ref{fig_compower2}, the power allocated entirely to the direct or IRS channel exceeds the partially allocated power under a certain invasion depth. Therefore, based on the changes in $d$ in Fig.~\ref{fig_dynamicpower1}, the power is entirely allocated to the direct LoS channel if $d$ is small, while the power is allocated to the IRS LoS channel as $d$ increases. For example, $P_{\rm{oc}}=P_{\rm{oc}}^{\rm{d}}$ with $\gamma=1$ when $d \leq 0.75$ mm, and $P_{\rm{oc}}=P_{\rm{oc}}^{\rm{i}}$ with $\gamma=0$ when $0.75 < d < 2.5$ mm. Similarly, as shown in Fig.~\ref{fig_dynamicpower2}, if the translation distance is small, i.e., $\Delta x \leq 0.5$ mm, $P_{\rm{oc}}=P_{\rm{oc}}^{\rm{d}}$ with $\gamma=1$, while $P_{\rm{oc}}=P_{\rm{oc}}^{\rm{i}}$ with $\gamma=0$ at larger translation distances, i.e., $\Delta x > 0.5$ mm. Finally, when the receiver rotates with an increasing angle, the power in the IRS-assisted channel is always greater than the power in the direct LoS channel and the combined power of both channels ($0 < \gamma < 1$) as shown in Fig.~\ref{fig_dynamicpower3}. Therefore, all transmitted power is allocated to the IRS-assisted channel under dynamic regulation, i.e., $P_{\rm{oc}}=P_{\rm{oc}}^{\rm{i}}$ with $\gamma=0$ at $\theta_{\rm{y}} \in [0^\circ, 50^\circ]$. Similarly, $\gamma$ can be dynamically adjusted to maximize power depending on different system parameters, channel parameters, and receiver conditions.

\subsection{Channel capacity}
According to \eqref{eq:SNR}, \eqref{eq:capacity} and the optimal communication power shown in Fig.\ref{fig_dynamicpower}, the $\mathrm{SNR}$ and spectral efficiency of the IRS-assisted RBC system under specific parameters can be calculated numerically. The variations of $\mathrm{SNR}$ and spectral efficiency $S_{\rm{E}}$ with different invasion depths, translation distances, and rotation angles are illustrated in Fig.\ref{Fig_com}. As depicted in Fig.\ref{Fig_com1}, when the transceiver is aligned, i.e., $\Delta x=0$ mm and $\theta_{\rm{y}}=0^{\circ}$, $\mathrm{SNR}$ initially experiences a slight decrease but then stabilizes, maintaining around $92.95$ dB.As $\Delta x$ and $\theta_{\rm{y}}$ increase, the $\mathrm{SNR}$ shows a significant decrease under the same invasion depth due to the reduction in output power, as shown in Figs.~\ref{fig_compower3} and \ref{fig_compower4}, with the invasion location at $z_{o}=0.5$. For example, $\mathrm{SNR}$ is approximately $87.60$ dB and $73.96$ dB at $\Delta x = 1$ mm and $2$ mm, respectively, and approximately $87.48$ dB and $79.31$ dB at $\theta_{\rm{y}}=10^{\circ}$ and $\theta_{\rm{y}}=20^{\circ}$, respectively, with $d > 1$ mm. Additionally, when the invasion depth is small ($d \leq 1$ mm), the $\mathrm{SNR}$ experiences a slight decrease but then remains constant despite the receiver misalignment.

Similarly, the spectral efficiency $S_{\rm{E}}$ follows the same trend as $\mathrm{SNR}$, as shown in Fig.~\ref{Fig_com2}. It exhibits a slight downward trend before stabilizing in both the aligned and misaligned cases. $S_{\rm{E}}$ remains at $10.09$ bps/Hz and $8.51$ bps/Hz at $\Delta x = 1$ mm and $2$ mm, respectively, and approximately $10.07$ bps/Hz and $9.13$ bps/Hz at $\theta_{\rm{y}}=10^{\circ}$ and $\theta_{\rm{y}}=20^{\circ}$, respectively, with $d > 1$ mm. The maximum spectral efficiency is approximately $11$ bps/Hz when the transceiver is aligned.

This numerical analysis of $\mathrm{SNR}$ and spectral efficiency reveals that the assistance of the IRS enables the RBC system to maintain a high level of operating efficiency despite external object invasion and transceiver misalignment.


\section{Conclusion}
In this paper, we propose an IRS-assisted resonant beam communications (RBC) system utilizing a frequency-doubling method to achieve highly efficient downlink communication in the presence of external object invasion and receiver misalignment. By leveraging near-field optical propagation, we establish a transmission model for both direct line-of-sight (LoS) and IRS-reflected channels, and then derive the output frequency-doubled beam power based on the transfer factor. To optimize the communication performance, we maximized the frequency-doubled power by dynamically adjusting the power-splitting ratio between the two channels, taking into account different obstruction depths and misalignment parameters. Finally, numerical evaluations demonstrated that with IRS assistance and power ratio regulation, the signal-to-noise ratio (SNR) and spectral efficiency can be maintained at optimal values, with the maximum spectral efficiency reaching approximately $11$ bps/Hz.

Additionally, several issues need to be addressed in future research, including: 1) analyzing uplink communication using resonant beams with IRS assistance; 2) evaluating communication performance metrics such as bit error rate (BER), Q factor, and modulation methods for RBC; and 3) optimizing system design to further enhance communication performance.

\ifCLASSOPTIONcaptionsoff
  \newpage
\fi

{\small
\bibliographystyle{IEEEtran}
\bibliography{references}
}


\end{document}